\documentclass[AMA,STIX1COL]{WileyNJD-v2}

\usepackage{tabularx,booktabs}
\usepackage{bbold}
\usepackage{color,colortbl}
\usepackage{bbm}
\usepackage{mathtools} 
\usepackage{etoolbox}
\usepackage{dsfont}
\usepackage{booktabs,caption,fixltx2e}
\usepackage{multirow}
\usepackage{graphicx}
\usepackage{tabularx,booktabs}
\usepackage{bbold}
\usepackage{subcaption}
\usepackage{color,colortbl}
\usepackage{bbm}
\usepackage{mathtools} 
\usepackage{etoolbox}
\usepackage{dsfont}
\usepackage{booktabs,caption,fixltx2e}
\usepackage[flushleft]{threeparttable}
\usepackage{multirow}
\usepackage{graphicx}
\usepackage{setspace}
\usepackage[utf8]{inputenc}
\usepackage{enumitem} 

\usepackage{comment}

\hypersetup{
    colorlinks,
    citecolor=black,
    filecolor=black,
    linkcolor=black,
    urlcolor=black
}

\AtBeginEnvironment{tabular}{\singlespacing}

\BeforeBeginEnvironment{verbatim}{\def\baselinestretch{1}}

\renewcommand{\baselinestretch}{1.4}
\def\singlespace{\def\baselinestretch{1}\@normalsize}

\usepackage{bm}

\catcode`@=11

\def\@evenhead{\vbox{\hbox to\textwidth{\footnotesize \hfill
 \hfill } }}
\usepackage{amsmath,graphicx,centernot}
\def\T{{\mbox{\rm\tiny T}}}

\newcommand{\CI}{\mathrel{\perp\mspace{-10mu}\perp}}

\def\T{{\mbox{\rm\tiny T}}}

\def\P{{\mbox{\rm\tiny P}}}

\def\m{{\mbox{\rm\tiny m}}}
\def\C{{\mbox{\rm\tiny C}}}
\def\B{{\mbox{\rm\tiny B}}}

\def\I{{\mbox{\rm\tiny I}}}
\def\W{{\mbox{\rm\tiny W}}}
\def\E{{\mbox{\rm\tiny E}}}
\def\L{{\mbox{\rm\tiny L}}}

\articletype{Research Article}%

\received{}
\revised{}
\accepted{}

\begin{document}

\title{Robust Causal Inference of Drug-drug Interactions}

\author[1,2,3,4]{Di Shu*}

\author[5]{Peisong Han}

\author[1,4]{Sean Hennessy}

\author[1,4]{Todd A Miano}


\address[1]{Department of Biostatistics, Epidemiology and Informatics, University of Pennsylvania Perelman School of Medicine, Philadelphia, PA, USA}

\address[2]{Department of Pediatrics,  University of Pennsylvania Perelman School of Medicine,
Philadelphia, PA, USA}

\address[3]{Center for Pediatric Clinical Effectiveness, Children’s Hospital of Philadelphia, Philadelphia, PA, USA}

\address[4]{Center for Real-world Effectiveness and Safety of Therapeutics, University of Pennsylvania Perelman School of Medicine, Philadelphia, PA, USA}

\address[5]{Department of Biostatistics, University of
Michigan, Ann Arbor, Michigan, USA}

\corres{*Di Shu, Department of Biostatistics, Epidemiology and Informatics, Perelman School of Medicine, University of Pennsylvania, 423 Guardian Drive Philadelphia, PA 19104-6021, USA\\
\email{di.shu@pennmedicine.upenn.edu}}

\abstract{There is growing interest in developing causal inference methods for multi-valued treatments with a focus on pairwise average  treatment effects. Here we focus on a clinically important, yet less-studied estimand: causal drug-drug interactions (DDIs), which quantifies the degree to which the causal effect of drug A is altered by the presence versus the absence of drug B. Confounding adjustment when studying the effects of DDIs can be accomplished via inverse probability of treatment weighting (IPTW), a standard approach originally developed for binary treatments and later generalized to multi-valued treatments. However, this approach generally results in biased results when the propensity score model is misspecified. Motivated by the need for more robust techniques, we propose two empirical likelihood-based weighting approaches that allow for specifying a set of propensity score models, with the second method balancing user-specified covariates directly, by incorporating additional, nonparametric constraints. The resulting estimators from both methods are consistent when the postulated set of propensity score models contains a correct one; this property has been termed multiple robustness. We then evaluate their finite sample performance through simulation. The results demonstrate that the proposed estimators outperform the standard IPTW method in terms of both robustness and efficiency. Finally, we apply the proposed methods to evaluate the impact of renin-angiotensin system inhibitors (RAS-I) on the comparative nephrotoxicity of nonsteroidal anti-inflammatory drugs (NSAID) and opioids, using data derived from electronic medical records from a large multi-hospital health system.}

\keywords{Causal inference, drug-drug interaction, multiple robustness, pharmacoepidemiology, propensity score weighting}


\maketitle

\section{Introduction}

Causal inference methods are increasingly used to assess   effects of treatments on health outcomes in both randomized clinical trials and observational studies. \citep{RosenbaumRubin1983, HernanRobins2019} Although a substantial amount of research has focused on methodology for examining the comparative safety and effectiveness  of one or more treatments of interest, considerably less attention has been given to  causal inference of interactions, where the target of inference is an interaction contrast such as a difference-in-difference or a ratio-of-ratios.  \citep{vanderweele2014}  Control of confounding in the context of an interaction can be more complex relative to standard treatment comparisons because there are two sets of confounders to consider, one for each exposure participating in the interaction. \citep{vanderweele2014,VanderWeele2009,HernanRobins2019} An area where interaction causal inference is particularly important is the study of drug-drug interactions (DDIs),\citep{HennessyFlockhart2012,Hennessy2016} which account for an estimated   13\% of adverse drug events in community-dwelling older adults. \citep{Qurwitz2003} DDI occurs when the pharmacologic effect of a drug is altered by the coadministration of a second drug, through either a pharmacokinetic (the body’s impact on the drug) and/or pharmacodynamic (the drug’s impact on the body) mechanism; the affected drug is called the object (or victim)
and the affecting drug is called the precipitant (or perpetrator). Causal inference methods for DDIs are useful for answering  the following key question:\citep{HennessyFlockhart2012}  To what degree does the DDI increase the absolute or relative risk  of an adverse drug event?

In principle, causal inference methods developed for multi-valued treatments can be applied to estimate the effects of DDIs, by noticing that a two-drug DDI fits into a 4-level treatment setting, when considering all four combinations of two drugs: i) object + precipitant, ii) object + control of precipitant, iii) control of object + precipitant, and iv) control of object + control of precipitant. In well-implemented randomized controlled trials with a $2\times 2$ factorial design, a causal DDI can be nonparametrically estimated. Specifically, the investigator first  estimates the mean of potential outcomes for each of the four treatment arms by the average of outcome data. A causal DDI interaction contrast, either a difference of differences or a ratio of ratios, is a function of the four averages and hence can be directly calculated from them. Such trials are, however, rarely if ever conducted. More commonly, observational studies are employed to examine DDI effects. In this context,   a widely used approach to confounding adjustment is through inverse probability of treatment weighting (IPTW), \citep{HorvitzThompson1952,RosenbaumRubin1983,Rosenbaum1987,LuncefordDavidian2004,SeamanWhite2013, HernanRobins2019} originally developed for binary treatments and later extended to multi-valued treatments. \citep{Imbens2000, Imai2004}  For each individual, the investigator  first estimates the weight by the reciprocal of the estimated probability of receiving the treatment level that this individual actually received, i.e., the estimated propensity score. Then, the investigator estimates the causal DDI using the four weighted averages of outcome data corresponding to four treatment levels, i.e., treating the weighted sample data as if they were collected from a  $2\times 2$ factorial trial.

One advantage of the IPTW approach is that the possibly complex outcome model is left completely unspecified. Instead, only a propensity score model needs to be specified. In many applications, it is more feasible to model the  probability of receiving  treatment than to model the outcome process, for example, in pharmacoepidemiologic studies with common treatments and rare adverse events. There is growing interest in developing methods for propensity score estimation with multi-valued treatments.\citep{Spreeuwenberg2010,McCaffrey2013,Imai2014,LiLi2019,Yoshida2019}  We take a different weighting approach based on empirical likelihood\citep{QinLawless1994, Owen2001,Chen2002,HanWang2013} to make use of a set of propensity score model simultaneously, with an option to directly balance  distributions of user-specified covariates. Our work was motivated by the need to increase the chance of correctly modeling the true treatment decision process, which is generally unknown in observational studies. It has long been recognized that model selection is one of the most challenging
steps of applied causal inference. For example, adding more variables into an existing model  can paradoxically increase bias and decrease efficiency; see Schisterman et al for a discussion of overadjustment bias and unnecessary adjustment. \citep{Schisterman2009}  Even with the help of useful guidelines, \citep{Mickey1989, Brookhart2006,Schneeweiss2009,vanderweele2019} it is often  difficult to develop a final, correct propensity score model.  Multiply robust estimation  \citep{HanWang2013, Han2014a, Han2014b, Chan2016, ChenHaziza2017, Shu2021} makes it possible not to rely on a single hopefully correct propensity score model, in the sense that its validity requires only one  model among all candidate models be correctly specified. Moreover, there is no need to identify the correct model.

In this paper,  we propose two multiply robust estimators of the causal DDI.  Both estimators  allow for specifying a set of propensity score  models, by setting up a set of constraints on these models when maximizing the empirical likelihood. The first method extends empirical likelihood weights in Shu et al\citep{Shu2021} to accommodate multi-valued treatments. As a major difference from the first method and the original empirical likelihood-based multiply robust method,\citep{HanWang2013, Han2014a, Han2014b} the second proposed method  imposes additional constraints to directly balance the distributions for user-specified covariates of interest. For each method, the resulting estimator of  the causal DDI  achieves statistical consistency when the postulated model set contains a correct one, thereby providing increased protection against model misspecification. To our knowledge, this work is the first to develop and evaluate multiply robust methods focusing on estimation of causal DDIs.

The remainder of this paper is organized as follows.  In Section  \ref{notation}, we introduce the notations and estimand. In Section  \ref{secMethod}, we first describe the standard IPTW method for causal inference of DDIs that uses a single propensity score model, then propose two multiply robust approaches making use of a set of propensity score models. In Sections \ref{secSimu1} and \ref{secSimu2}, we conduct simulations to assess the finite sample performance of the proposed methods compared with the IPTW method, with and without including a correctly specified propensity score model in the set of postulated
models, respectively. In Section \ref{secAppl},  we  apply the proposed methods  to estimate the causal interaction between nonsteroidal anti-inflammatory drugs (NSAID) and renin-angiotensin system inhibitors (RAS-I) with regard to acute kidney injury (AKI)
outcomes in hospitalized patients. Specifically, we contrast the difference in AKI risk between NSAID versus oxycodone in patients treated with RAS-I and the difference in AKI risk between NSAID versus oxycodone in patients treated with amlodipine (an active comparator for RAS-I treatment).  We  conclude the paper with a discussion in Section \ref{secDiscuss}.

\section{Data Structure and Estimand}\label{notation}

We begin by introducing the notation.  In our  two-drug DDI context,  let $A$ be the observed treatment indicator for the affected drug (also called object or victim) such that $A=1$ if treated and $A=0$ if untreated; let $B$ be the observed treatment indicator for the affecting drug (also called the precipitant or perpetrator)  such that  $B=1$ if treated and $B=0$ if untreated. Let $\bm X$ be a vector of measured baseline covariates and $Y$  the observed outcome.

Following the potential outcome framework in causal inference,\citep{Rubin1974,Holland1986} let $Y_{ab}$ represent the potential outcome for an individual that would have been observed had we set the treatment levels to $a$ and $b$  for $a=1$ (treated with the object) or $0$ (untreated with the object) and $b=1$ (treated with the precipitant) or $0$ (untreated with  the precipitant). Each individual has four potential outcomes, $Y_{11}$, $Y_{10}$, $Y_{01}$ and $Y_{00}$, but only the one under the actually observed treatment status can be observed. That is, $Y=Y_{ab}$ when $A=a$ and $B=b$; this is the consistency assumption in causal inference. \citep{HernanRobins2019}

Suppose we have an independent and identically distributed (i.i.d.) sample of size $n$. The observed data are $(Y_i,  \bm X_i, A_i, B_i)$ for the $i$th individual, where  $i=1,\ldots, n$. We aim to use the observed data to estimate the causal effect of DDI, $\theta$, defined in the following general form: 

\begin{equation}
\theta=[f\{E(Y_{11})\}-f\{E(Y_{01})\}]-[f\{E(Y_{10})\}-f\{E(Y_{00})\}],
\label{ddi}
\end{equation}
where $f(\cdot)$ is a known monotone link function. 

When $f(\cdot)$ is an identity link function, (\ref{ddi}) reduces to a  measure of interaction on the additive scale. That is, $\theta=\{E(Y_{11})-E(Y_{01})\}-\{E(Y_{10})-E(Y_{00})\}$, where $E(Y_{11})-E(Y_{01})$ represents the average causal effect of the object  had the entire population been treated with the precipitant, and $E(Y_{10})-E(Y_{00})$ represents the average causal effect of the object   had the entire population been untreated with the precipitant. A non-zero $\theta$ suggests that the  presence of the precipitant alters the magnitude of the effect of the object. In other words, the DDI does exist. Equivalently, $\theta$ can be written as
$
\{E(Y_{11})-E(Y_{10})\}-\{E(Y_{01})-E(Y_{00})\},
$
representing the difference between the average causal effect of precipitant  had the entire population been treated with the object   and the average causal effect of precipitant  had the entire population been untreated with the object. With binary $Y$, $\theta$ can be interpreted as the causal risk difference comparing  object and its control in the presence of the  precipitant,  minus the causal risk difference comparing  object and its control in the absence of the  precipitant. As noted by VanderWeele, \citep{vanderweele2014}  the interaction is sometimes said to be positive (or super-additive) if $\theta>0$ and  negative (or  sub-additive) if $\theta<0$. If $\theta=0$, then there is no interaction on the additive scale for risk  differences.

When $f(\cdot)$ is a log link function and that the outcome is binary, (\ref{ddi}) reduces to a  measure of interaction on the multiplicative scale for risk ratios. \citep{vanderweele2014} In this case, $\exp(\theta)$ represents the ratio of two causal risk ratios, one risk ratio comparing the object and its control in the presence of the precipitant and the other risk ratio comparing the object and its control in the absence of the precipitant.  Similar to the DDI on the additive scale, this multiplicative interaction does not change when exchanging the order of two drugs.  It is said to be positive if $\exp(\theta)>1$ and negative if  $\exp(\theta)<1$.  If $\exp(\theta)=1$, i.e., $\theta=0$, then there is no interaction on the multiplicative scale for risk ratios.

\section{Causal Inference of Drug-drug Interactions Using the Propensity Score} \label{secMethod}

To consistently estimate the causal DDI measure $\theta$ in (\ref{ddi}), it suffices to consistently estimate $E(Y_{ab})$ for each of the four treatment combinations: i) $a=1, b=1$, ii) $a=0, b=1$, iii) $a=1, b=0$, and iv) $a=0, b=0$. In randomized controlled trials with a $2\times 2$ factorial design,  $E(Y_{ab})$  can be nonparametrically estimated  by  the average  outcome across individuals exposed to treatment level $A=a$ and $B=b$.  In observational studies, this can be accomplished using adjusted averages that account for measured confounding. Here, we focus on confounding adjustment via weighting.

Specifically, we consider a general form of causal DDI estimator:
\begin{equation}
\widehat\theta=[f\{\widehat E(Y_{11})\}-f\{\widehat E(Y_{01})\}]-[f\{\widehat E(Y_{10})\}-f\{\widehat E(Y_{00})\}],
\label{est}
\end{equation}
where $\widehat E(Y_{ab})$ is the weighting-based estimator of $E(Y_{ab})$  given by
\begin{equation}
\widehat E(Y_{ab})=\sum_{i: A_i=a, B_i=b}\widehat w{_i}Y_i
\label{estcontinue}
\end{equation}
for $a, b=1, 0$.

To calculate $\widehat w_i$ in (\ref{estcontinue}), we describe three propensity score-based weighting methods in next sections: the IPTW approach,  the empirical likelihood approach, and a modified empirical likelihood approach with covariate balancing.

\subsection{The IPTW Approach Using A Single Propensity Score Model}\label{secIPTW}

We apply  IPTW for multi-valued  treatments, originally developed by Imbens \citep{Imbens2000} as an extension of the propensity score methodology for binary treatments,  to calculate the weights for estimating the causal DDI. The  reason for taking this approach is that a two-drug DDI can be understood as a multi-valued treatment setting, with a composite  treatment (combining the object and precipitant)  having four levels:  $(a,b)=(1,1), (0,1), (1,0)$ and $(0,0)$.

Just like in IPTW estimation with binary treatments, causal inference assumptions (including  consistency, exchangeability, and positivity \citep{HernanRobins2019})  are required for valid IPTW estimation with multi-valued treatments. \citep{Imbens2000,McCaffrey2013} The consistency assumption means the observed outcome equals the potential outcome under the actually observed treatment.  The exchangeability assumption, termed weak unconfoundedness by Imbens,  is satisfied   if $T(a,b) \CI  Y_{ab}  |  \bm X$,  where $T_i(a,b)=I(A_i=a,B_i=b)$. That is,  adjusting for $\bm X$ is sufficient to eliminate confounding bias; thus exchangeability is often also called an assumption of no unmeasured confounding. Here,  the role of two exposures are symmetric in an interaction, and both exposures (or the  composite  exposure) are required to be unconfounded. These features distinguish interaction from  the concept of effect modification.  \citep{VanderWeele2009,HernanRobins2019}
The positivity assumption states that each individual has a positive probability of receiving each treatment. Formally, $0<P(A=a,B=b|\bm X)<1$ for all $\bm X$ and for $a, b=1$ or 0.

Provided that these standard causal inference assumptions are satisfied, propensity score weighting effectively eliminates confounding bias because the weighted data in theory emulate data that would have been collected from a $2\times 2$ factorial trial.

Since the true treatment decision process is often unknown in observational studies, the propensity score $e_{ab}(\bm X)=P(A=a, B=b|\bm X)$ needs to be estimated from the data. A  modeling strategy is to fit a parametric propensity score model for $e_{ab}(\bm X)$ that relates the treatment variable to baseline covariates  $\bm X$.  Let $\widehat P(A_i=a, B_i=b|\bm X_i)$ denote the resulting estimate of the propensity score $P(A_i=a, B_i=b|\bm X_i)$ for the $i$th individual. Then the IPTW weights are given by
\begin{equation}
\widehat w_{\I\P\T\W,i}=\dfrac{1}{\widehat P(A_i=a, B_i=b|\bm X_i)}\bigg / \sum_{i: A_i=a, B_i=b}\dfrac{1}{\widehat P(A_i=a, B_i=b|\bm X_i)} \,\,\text{for $i$ with $A_i=a, B_i=b$}.
\label{ipW}
\end{equation}

Substituting (\ref{ipW}) into (\ref{estcontinue}) gives a commonly-used ratio estimator in the IPTW estimation literature: \citep{LuncefordDavidian2004}
\[
\widehat E(Y_{ab})=\left\{\sum_{i: A_i=a, B_i=b} \dfrac{Y_i}{\widehat P(A_i=a, B_i=b|\bm X_i)} \right\}\bigg / \left\{\sum_{i: A_i=a, B_i=b} \dfrac{1}{\widehat P(A_i=a, B_i=b|\bm X_i)} \right\}.
\]
This estimator can be regarded  as a Hájek-type  estimator in the sampling literature.\citep{Hajek1971} It was found to have higher efficiency for estimating the average causal treatment effect than the Horvitz-Thompson-type counterpart. \citep{LuncefordDavidian2004}

In practice,  one single  propensity score model is specified and fitted to the data, and multinomial logistic regression has been commonly employed with standard software packages such as the R  function {\it multinom}. This model describes $e_{ab}(\bm X)$ using three vectors of parameters $\bm\gamma_{11}$, $\bm\gamma_{01}$ and $\bm\gamma_{10}$: 
\begin{equation*}
P(A_i=a, B_i=b|\bm X_i)=\dfrac{\exp(\bm\gamma_{ab}^\T\bm X_i)}{1+\exp(\bm\gamma_{11}^\T\bm X_i)+\exp(\bm\gamma_{01}^\T\bm X_i)+\exp(\bm\gamma_{10}^\T\bm X_i)}\quad\text{for $(a,b)=(1,1), (0,1)$ or $(1,0)$}
\end{equation*}
and
\begin{equation*}
P(A_i=a, B_i=b|\bm X_i)=\dfrac{1}{1+\exp(\bm\gamma_{11}^\T\bm X_i)+\exp(\bm\gamma_{01}^\T\bm X_i)+\exp(\bm\gamma_{10}^\T\bm X_i)}\quad\text{for $(a,b)=(0,0)$},
\end{equation*}
where treatment $(0,0)$ is considered the reference level; such choice of reference level is arbitrary. Write $\bm\gamma=(\bm\gamma_{11}^\T, \bm\gamma_{01}^\T, \bm\gamma_{10}^\T)^\T$. Fitting the model gives an estimate of $\bm\gamma$ and hence estimates of $e_{ab}(\bm X)$
for $a,b=1$ or $0$.

It is also possible to estimate $P(A_i=a, B_i=b|\bm X_i)$ by multiplying estimates for probabilities $P(A_i=a |\bm X_i)$ and $P(B_i=b|A_i=a, \bm X_i)$. This can be accomplished by fitting two binary logistic models, one relating $A$ to $\bm X$ and the other relating $B$ to  $A$ and $\bm X$.

Substituting $\widehat w_i=\widehat w_{\I\P\T\W, i}$ into  (\ref{estcontinue}) and then applying  (\ref{est}) gives the causal DDI estimator based on IPTW, denoted as $\widehat\theta_{\I\P\T\W}$. The consistency of estimator  $\widehat\theta_{\I\P\T\W}$   requires a correctly specified propensity score model. Misspecifying the  propensity score model generally leads to  biased results.

\subsection{The Empirical Likelihood Approach Using Multiple Propensity Score Models}  \label{secProp1}

Placing the development in biased sampling context (e.g., the likelihood function contributed by treated individuals is only a biased version of the likelihood function that would have been obtained had all individuals been treated), empirical likelihood  theory  \citep{QinLawless1994, Owen2001}  provides a different confounding adjustment approach from IPTW. \citep[][Ch.19]{Qin2017}    To correct biased samples, Qin and Zhang  \cite{QinZhang2007}  added two constrains to the biased likelihood to be maximized, with one constraint on the propensity score and the other on a function of covariates. Han and Wang \cite{HanWang2013} and Han  \cite{Han2014a, Han2014b, Han2016} developed an extended approach to incorporating a set of constraints on the propensity score model as well as a set of  constraints on the outcome model. 

With binary treatments, Shu et al \citep{Shu2021} adapted this methodology by using propensity score constraints only. They did not use any outcome model constraint in order to get around the noncollapsibility issue  for  conditional and marginal Cox models. One practical advantage of their approach is, the resulting weights  are generally applicable to other outcome types. Here, we first generalize this strategy to  multi-valued treatment case and then use the resulting weights to estimate the causal DDI. Our method allows the investigator to postulate a set of propensity score models simultaneously. The resulting DDI estimator is statistically consistent when this model set contains a correct one. We describe the estimation procedures below.

\subsubsection{Postulating A Set of Propensity Score Models}
Instead of using just one parametric model for the propensity score $e_{ab}(\bm X)=P(A=a, B=b|\bm X)$, we postulate a set of $J$  models  to increase the chance of modeling $e_{ab}(\bm X)$ correctly,   where $J$ is a user-specified, arbitrary positive integer. Let  $\mathcal{E}=\{e_{ab}^j(\bm\gamma^j; \bm X): j=1,\ldots,J\}$ denote this set of  $J$ postulated propensity score models, where $\bm\gamma^j$ is the vector of  parameters for the $j$th  model.

For $j=1,\ldots, J$, fitting the $j$th propensity score model gives an estimator of ${\bm\gamma}^j$, denoted as  $\widehat{\bm\gamma}^j$. We define $\widehat{\bm\gamma}=(\widehat{\bm\gamma}^{1\T},\ldots,\widehat{\bm\gamma}^{J\T})^\T$.  For $a, b=1$ or 0, we define $\widehat\mu_{ab}^j=n^{-1}\sum_{i=1}^n e_{ab}^j(\widehat{\bm\gamma}^j; \bm X_i)$ for $j=1,\ldots,J$ and 
\begin{equation}
\widehat g_{ab}(\widehat{\bm\gamma}; i)=(e_{ab}^1(\widehat{\bm\gamma}^1; \bm X_i)-\widehat\mu_{ab}^1,\ldots, e_{ab}^J(\widehat{\bm\gamma}^J; \bm X_i)-\widehat\mu_{ab}^J)^\T 
 \label{gfunction}
\end{equation}
for $i=1,\ldots,n$. Note that the superscript $j$ is needed for $\widehat\mu_{ab}^j$ because $n^{-1}\sum_{i=1}^n e_{ab}^j(\widehat{\bm\gamma}^j; \bm X_i)$ depends on the $j$th model and may not be the marginal probability $P(A=a,B=b)$. Consider an example with binary treatments. Suppose the true propensity score model is a standard logistic model with intercept whereas the $j$th postulated model is a logistic model without intercept. In this case, $\widehat\mu_{ab}^j$ can be different from the marginal treatment prevalence; interestingly, any logistic model with intercept, even if misspecified, would ensure that $\widehat\mu_{ab}^j$ equals the treatment prevalence through solving the estimating equation for the intercept parameter.

\subsubsection{Empirical Likelihood Weights}

For  individuals whose observed treatment level is $(a,b)$, consider   a constrained optimization problem with respect to weights $\{w_i\}$ jointly:
\begin{equation*}
{\text{max}}\prod_{i: A_i=a, B_i=b} w_i 
\end{equation*}
subject to constraints

\begin{equation}
w_i\ge 0, \sum_{i: A_i=a, B_i=b}  w_i=1, \quad \text{and}  \quad  \sum_{i: A_i=a, B_i=b} w_i\widehat g_{ab}(\widehat{\bm\gamma}; i)=\bm 0 .
 \label{emcons1}
\end{equation}
Here  $w_i$ represents the empirical probability of $(Y_i,X_i)$ conditional on $A_i=a,B_i=b$.\citep{HanWang2013} Solving this optimization problem amounts to finding the most plausible values of  $\{w_i\}$ such that the resulting empirical likelihood (i.e., product of empirical probabilities) is maximized.

The third constraint in (\ref{emcons1}) is a weighted summation using data contributed by individuals with observed treatment level $(a,b)$ only.  By definition, $\sum_{i=1}^n \widehat g_{ab}(\widehat{\bm\gamma}; i)= \bm 0$,  indicating that the unweighted summation using the entire sample also equals zero. Thus, the  constraint $ \sum_{i: A_i=a, B_i=b} w_i\widehat g_{ab}(\widehat{\bm\gamma}; i)=\bm 0$ produces weights to re-balance the  group of treatment level $(a,b)$, an originally biased sample of the entire sample. For  binary treatments, constraints  (\ref{emcons1}) reduce to that considered by Shu et al, \citep{Shu2021} as a modification to the original constraints in Han and Wang \cite{HanWang2013} and Han, \cite{Han2014a, Han2014b, Han2016} by not using constraints on conditional outcome models.

Solving the above constrained optimization gives the empirical likelihood weights   for individuals $\{i: A_i=a, B_i=b\}$:
\begin{equation}
\widehat w_{\E\L,i}=\left\{\dfrac{1}{1+\widehat{\bm\rho}_{ab}^\T \widehat g_{ab}(\widehat{\bm\gamma}; i) }\right\}\bigg/\sum_{i: A_i=a, B_i=b}\left\{\dfrac{1}{1+\widehat{\bm\rho}_{ab}^\T \widehat g_{ab}(\widehat{\bm\gamma}; i) }\right\} \,\,\text{for $i$ with $A_i=a, B_i=b$}
 \label{ELweights}
\end{equation}
where for $a, b=1$ or 0,
$\widehat{\bm\rho}_{ab}$ is a $J\times 1$ vector that satisfies
\[
\sum_{i: A_i=a, B_i=b}\dfrac{ \widehat g_{ab}(\widehat{\bm\gamma}; i)}{1+\widehat{\bm\rho}_{ab}^\T  \widehat g_{ab}(\widehat{\bm\gamma}; i)}=\bm 0 
\]
and can be obtained via the convex minimization method of Han. \citep{Han2014a}

\subsubsection{Proposed Weighted Estimation Method}\label{proof1}

We propose to estimate the causal DDI by substituting the empirical likelihood weights $\widehat w_i=\widehat w_{\E\L, i}$   into  (\ref{estcontinue}) and then applying  (\ref{est}). Denote the resulting estimator  as $\widehat\theta_{\E\L}$.  Applying existing proofs \cite{HanWang2013,Shu2021} to each treatment level $(a,b)$ reveals multiple robustness of $\widehat\theta_{\E\L}$. That is, $\widehat\theta_{\E\L}$ consistently estimates the causal DDI $\theta$ in (\ref{ddi}), if the postulated set of propensity score models $\mathcal{E}=\{e_{ab}^j(\bm\gamma^j; \bm X): j=1,\ldots,J\}$ contains a correctly specified one. Specifically, the proposed weights are asymptotically equivalent to the IPTW weights obtained by fitting a correctly specified propensity score model. Therefore, weights $\widehat w_{\E\L, i}$ are expected to eliminate confounding bias and  lead to a consistent estimator of $\theta$ if the postulated set of propensity score models contains the true model, just like the  IPTW weights from a correct propensity score model. The empirical likelihood approach can be understood as a way to automatically select a correct model from postulated models.

\subsection{Modified Empirical Likelihood Weights with  Covariate Balancing}  \label{secProp2}

In practice, it is sometimes desirable to assure satisfactory covariate balance of certain user-specified, important covariates. However, using a good propensity score model (or even the true model)  does not always produce optimal balances given that chance imbalances may still occur in finite samples.  While our empirical likelihood weighting approach in Section \ref{secProp1} reduces model dependence by increasing the number of postulated models (and hence the chance of correctly estimating propensity scores), it does not directly, nonparametrically enforce the covariate balance.

In this section, we  propose a modified empirical likelihood weighting approach, in order to achieve  covariate balance for a set of user-specified covariates in finite samples. Let $X_{s_1},\ldots, X_{s_P}$ denote this set of $P$ covariates. For example, if we aim to balance the distributions of age and chronic kidney disease, then $P=2$.  To incorporate this need into estimating procedures described in Section \ref{secProp1}, we now define a vector: $(X_{s_1i}-\bar X_{s_1},\ldots , X_{s_Pi}-\bar X_{s_P})^\T$, where $X_{s_l i}$ denotes  covariate $X_{s_l}$ for the $i$th individual, and $\bar X_{s_l}=\dfrac{1}{n}\sum_{i=1}^n X_{s_li}$ denotes the unweighted mean of the covariate $X_{s_l}$ for the pooled sample across all four treatment combinations, for $l=1,\ldots, P$.

We add this vector into  (\ref{gfunction}) and obtain an augmented version of function $g_{ab}(\cdot)$:
 \begin{equation}
\widehat g^+_{ab}(\widehat{\bm\gamma}; i)=(e_{ab}^1(\widehat{\bm\gamma}^1; \bm X_i)-\widehat\mu_{ab}^1,\ldots, e_{ab}^J(\widehat{\bm\gamma}^J; \bm X_i)-\widehat\mu_{ab}^J, X_{s_1i}-\bar X_{s_1},\ldots , X_{s_Pi}-\bar X_{s_P})^\T .
 \label{gplusfunction}
\end{equation}

 For  individuals $\{i: A_i=a, B_i=b\}$, consider  empirical likelihood weights defined through the following constrained optimization problem with respect to weights $\{w_i\}$ jointly:
\begin{equation*}
{\text{max}}\prod_{i: A_i=a, B_i=b} w_i 
\end{equation*}
subject to constraints

\begin{equation}
w_i\ge 0, \sum_{i: A_i=a, B_i=b}  w_i=1, \quad \text{and}  \quad  \sum_{i: A_i=a, B_i=b} w_i\widehat g^+_{ab}(\widehat{\bm\gamma}; i)=\bm 0 .
 \label{emcons2}
\end{equation}

The resulting empirical likelihood weights are
\begin{equation}
\widehat w_{\m\E\L\C\B,i}=\left\{\dfrac{1}{1+\widehat{\bm\eta}_{ab}^\T \widehat g^+_{ab}(\widehat{\bm\gamma}; i) }\right\}\bigg/\sum_{i: A_i=a, B_i=b}\left\{\dfrac{1}{1+\widehat{\bm\eta}_{ab}^\T \widehat g^+_{ab}(\widehat{\bm\gamma}; i) }\right\}\,\,\text{for $i$ with $A_i=a, B_i=b$}
 \label{ELCBweights}
\end{equation}
where for $a, b=1$ or 0, 
$\widehat{\bm\eta}_{ab}$ is a $(J+P)\times 1$ vector  that satisfies
\[
\sum_{i: A_i=a, B_i=b}\dfrac{ \widehat g^+_{ab}(\widehat{\bm\gamma}; i)}{1+\widehat{\bm\eta}_{ab}^\T  \widehat g^+_{ab}(\widehat{\bm\gamma}; i)}=\bm 0 
\]
and can be obtained via convex minimization. \citep{Han2014a}

We propose to estimate the causal DDI by substituting the modified empirical likelihood weights  $\widehat w_i=\widehat w_{\m\E\L\C\B, i}$   into  (\ref{estcontinue}) and then applying  (\ref{est}). Denote the resulting estimator  as $\widehat\theta_{\m\E\L\C\B}$. Similar to justification in Section \ref{proof1},  multiple robustness of $\widehat\theta_{\m\E\L\C\B}$ is guaranteed as long as $\mathcal{E}=\{e_{ab}^j(\bm\gamma^j; \bm X): j=1,\ldots,J\}$   contains a correctly specified propensity score model, because under this condition the proposed weights are asymptotically equivalent to the IPTW weights obtained from a correct model. In other words, including additional  mean-centered covariates  helps achieve covariate balance without breaking multiple robustness.

\subsection{Estimation of Variance and $95\%$ Confidence Interval}

We apply the nonparametric bootstrap method \citep{EfronTibshirani1993} for variance estimation of the IPTW estimator $\widehat\theta_{\I\P\T\W}$ and the proposed empirical likelihood-based estimators $\widehat\theta_{\E\L}$ and $\widehat\theta_{\m\E\L\C\B}$.  Unlike the robust variance estimation which tends to be conservative, the bootstrap method provides more efficient inference by taking the uncertainty in weight estimation  into account.\citep{LuncefordDavidian2004, Austin2016, Shu2020} In the binary treatment setting of marginal hazard ratio estimation, prior simulations reported satisfactory performance of bootstrapping using $200$ replicates, for both the IPTW and empirical likelihood methods. \citep{Austin2016, Shu2021}

We describe the bootstrap procedures using $\widehat\theta_{\m\E\L\C\B}$ as an example. We first resample the observed data $(Y_i,  \bm X_i, A_i, B_i)$, $i=1,\ldots, n$ at the individual level with replacement  $R$ times to construct $R$ bootstrap samples each having the same sample size $n$ as the original data, where $R$ is a user-specified positive integer (say, $R=200$). For $r=1,\ldots, R$,  let  $\widehat\theta_{\m\E\L\C\B, r}$  denote the estimated causal DDI in the $r$th bootstrap sample using the modified empirical likelihood approach with  covariate balancing; here the weights are estimated using the same $r$th bootstrap sample rather than the original data, which accounts for uncertainty in estimation of weights. Then, the bootstrap variance estimators for  $\widehat\theta_{\m\E\L\C\B}$  is given by
\begin{equation*}
\widehat{var}(\widehat\theta_{\m\E\L\C\B})=\dfrac{1}{R-1}\sum_{r=1}^R\left(\widehat\theta_{\m\E\L\C\B, r}-\dfrac{1}{R}\sum_{r=1}^R \widehat\theta_{\m\E\L\C\B, r}\right)^2 .
\label{varBoot3}
\end{equation*}
A normality-based 95\% confidence interval for   $\widehat\theta_{\m\E\L\C\B}$ is  $\widehat\theta_{\m\E\L\C\B} \pm 1.96\cdot\sqrt{\widehat{var}(\widehat\theta_{\m\E\L\C\B})}$. Variance and confidence interval estimation for $\widehat\theta_{\I\P\T\W}$ and $\widehat\theta_{\E\L}$ can be obtained in the same manner.

\subsection{Balance Diagnosis in Two-drug Interaction Context}

Balance diagnosis  is an important component in a propensity score weighting-based  analysis. McCaffrey et al \citep{McCaffrey2013} described useful balance metrics  for multiple treatments  when using IPTW. They  can be adapted to our two-drug DDI context and to other weighting methods. Here we focus on metrics using standardized mean bias.

For each treatment level $(a,b)$ and each of the three methods (with $\widehat w_i$ given by $\widehat w_{\I\P\T\W, i}$, $\widehat w_{\E\L,i}$ and $\widehat w_{\m\E\L\C\B,i}$), balance for covariate $X_k$ can be summarized by the following absolute  standardized difference (also called population standardized bias\citep{McCaffrey2013}):   
\[
\text{PSB}_{ab,k}=|\bar X_{k,ab}-\bar X_{k}|/\widehat\sigma_{k},
\]
where 
\[
\bar X_{k, ab}=\sum_{i: A_i=a,B_i=b}\widehat w_i X_{ki}
\]
is the  weighted mean of this covariate within treatment group $(a,b)$, and $\bar X_{k}$ and $\widehat\sigma_{k}$ denote the unweighted
mean and standard deviation of this covariate in the pooled sample across all four treatment levels $(a,b)=(1,1), (0,1), (1,0)$ and $(0,0)$. 

A useful overall summary of balance across four treatment levels is 
\[
\text{PSB}_{k} = \max_{ab: a, b=1, 0} \text{PSB}_{ab,k}.
\]
The investigator can then report $\text{PSB}_{k}$ across covariates  before and after weighting. An absolute standardized difference larger than 0.2 was considered problematic. \citep{McCaffrey2013}

Notably,  by (\ref{gplusfunction}) and (\ref{emcons2}), if a covariate $X_k$ is  in the  set  $\{X_{s_1}\ldots, X_{s_P}\}$ for which  additional covariate balancing constraints are used, 
$
 \sum_{i: A_i=a, B_i=b} \widehat w_{\m\E\L\C\B,i} (X_{ki}-\bar X_{k})=\bm 0,
$
 leading to
 $
 \sum_{i: A_i=a, B_i=b} \widehat w_{\m\E\L\C\B,i} X_{ki}= \bar X_{k}
 $
 and  hence $\bar X_{k,ab}=\bar X_{k}$ for the weighted sample based on the modified empirical likelihood approach. These results reveal the relation between  the balance constrains  with  balance metrics $\text{PSB}_{ab,k}$ and $\text{PSB}_{k}$: the proposed modified empirical likelihood  approach aims to optimize  balance metrics $\text{PSB}_{ab,k}$ and $\text{PSB}_{k}$ for those user-specified covariates.

Above balance metrics compare each treatment level with the unweighted, pooled sample across four treatment levels.
It is also possible to check for covariate balance  for each of the six pairwise comparisons: $(1,1)$ vs $(1, 0)$, $(1,1)$ vs $(0, 1)$, $(1,1)$ vs $(0, 0)$, $(1,0)$ vs $(0, 1)$, $(1,0)$ vs $(0, 0)$ and $(0,1)$ vs $(0, 0)$. Within each  comparison, the  standardized difference measure across covariates  can be evaluated and a cutoff point at $0.1$ is  commonly used. \citep{AustinStuart2015}

\section{Simulation Studies: When the Postulated Set of Propensity Score Models Contains a Correct One}\label{secSimu1}

We conducted simulation studies to evaluate the finite sample performance of the proposed multiply robust estimators  $\widehat\theta_{\E\L}$ and $\widehat\theta_{\m\E\L\C\B}$ in comparison  to the standard IPTW estimator $\widehat\theta_{\I\P\T\W}$.

\subsection{Data Generating Process: The True Model}

For each individual, we first simulated a vector of covariates $\bm X=(X^{(1)}, X^{(2)}, X^{(3)}, X^{(4)}, X^{(5)})^\T$.  These covariates were independently simulated  with $X^{(1)}$ following a
Bernoulli distribution with  probability 0.2; $X^{(2)}$ following a Bernoulli distribution with  probability 0.4; $X^{(3)}$  following the standard normal distribution; $X^{(4)}$  following a uniform distribution ranging from -0.5 to 0.5; and $X^{(5)}$  following a unit exponential distribution.

We then simulated the treatment indicators $A$ and $B$ from the following multinomial logistic propensity score model:

\begin{equation}
P(A=a, B=b|\bm X)=\dfrac{Q_{ab}}{1+Q_{11}+Q_{01}+Q_{10}}\quad\text{for $(a,b)=(1,1), (0,1)$ or $(1,0)$},
\label{trtMsimu}
\end{equation}
where
\[
Q_{11}=\exp\left[\gamma_0+0.7+0.4X^{(1)}+0.2X^{(2)}-0.2X^{(3)}-0.4\exp(X^{(4)})-0.4\exp(X^{(5)})+0.2\{X^{(1)}X^{(3)}+X^{(2)}\exp(X^{(4)})\}\right],
\]
\[
Q_{01}=\exp\left[\gamma_0+0.6+0.2X^{(1)}+0.6X^{(2)}-0.4X^{(3)}-0.6\exp(X^{(4)})-0.2\exp(X^{(5)})+0.2\{X^{(1)}X^{(3)}+X^{(2)}\exp(X^{(4)})\}\right],
\]
\[
Q_{10}=\exp\left[\gamma_0+0.5+0.6X^{(1)}+0.4X^{(2)}-0.2X^{(3)}-0.2\exp(X^{(4)})-0.2\exp(X^{(5)})+0.4\{X^{(1)}X^{(3)}+X^{(2)}\exp(X^{(4)})\}\right],
\]
and the parameter $\gamma_0$ was selected to yield desired prevalence of treatment  level $(0,0)$ of  10\%, 20\%, 30\%, 40\% or 50\%. By definition,  
\[
P(A=a, B=b|\bm X)=\dfrac{1}{1+Q_{11}+Q_{01}+Q_{10}}\quad\text{for $(a,b)=(0,0)$}.
\]

Next, we generated the binary outcome $Y$ from the following  outcome model:
\begin{eqnarray*}
&&\text{logit}\,\, P(Y=1|A,B,\bm X)= 0.5-0.1 A-0.2 B-\xi AB  \nonumber \\
&+& 0.4 \left\{X^{(1)}+X^{(2)}-X^{(3)}+\exp(X^{(4)})-\exp(X^{(5)})+X^{(1)}X^{(3)}+X^{(2)}\exp(X^{(4)})\right\}\{0.5+\xi (0.2A+0.1B+A B)\},
\end{eqnarray*}
where $\xi=0.5, 1$ or 2, corresponding to  a resulting causal DDI of -0.107, -0.170 or -0.231.

With above set-up, we assume covariates $X^{(1)}$ to $X^{(5)}$ (and their functions) are sufficient for confounding control.

\subsection{Propensity Score Models Used to Analyze Simulated Data}

We considered   a set of four propensity score models, denoted as $\mathcal{E}=\{e_{ab}^j(\bm\gamma^j; \bm X): j=1,\ldots,4\}$, to  analyze the simulated data:

\begin{equation}
\text{Multinomial logistic}\,\, P(A=a, B=b|\bm X) \sim X^{(1)} + X^{(2)}\,\,\text{with parameters ${\bm\gamma}^1$},
\label{wrongPS1}
\end{equation}
\begin{equation}
\text{Multinomial logistic}\,\, P(A=a, B=b|\bm X) \sim X^{(1)} + X^{(2)} + X^{(3)} + X^{(4)} + X^{(5)}\,\,\text{with parameters ${\bm\gamma}^2$},
\label{wrongPS2}
\end{equation}
\begin{equation}
\text{Multinomial logistic}\,\, P(A=a, B=b|\bm X) \sim X^{(1)} + X^{(3)} + X^{(4)} \,\,\text{with parameters ${\bm\gamma}^3$},
\label{wrongPS3}
\end{equation}
\noindent and
\begin{eqnarray}
\text{Multinomial logistic}\,\, P(A=a, B=b|\bm X)  & \sim & X^{(1)} + X^{(2)} + X^{(3)} + \exp(X^{(4)}) + \exp(X^{(5)}) \nonumber\\
&& + \{X^{(1)}X^{(3)}+X^{(2)}\exp(X^{(4)})\}
\,\,\text{with parameters ${\bm\gamma}^4$},
\label{correctPS}
\end{eqnarray}
where  ${\bm\gamma}^j$ was the  vector of the parameters in the $j$th propensity score model, for $j=1,\ldots, 4$.

Taking model (\ref{wrongPS1}) as an example, with treatment $(0,0)$ considered the reference level, this model assumes
\begin{equation*}
P(A=a, B=b|\bm X)=\dfrac{\exp\{\bm\gamma_{ab}^{1\T}(X^{(1)}, X^{(2)})^\T\}}{1+\exp\{\bm\gamma_{11}^{1\T}(X^{(1)}, X^{(2)})^\T\}+\exp\{\bm\gamma_{01}^{1\T}(X^{(1)}, X^{(2)})^\T\}+\exp\{\bm\gamma_{10}^{1\T}(X^{(1)}, X^{(2)})^\T\}}
\end{equation*}
for $(a,b)=(1,1), (0,1)$ or $(1,0)$ where ${\bm\gamma}^1=(\bm\gamma_{11}^{1\T}, \bm\gamma_{01}^{1\T}, \bm\gamma_{10}^\T)^{\T}$.

Among these four postulated propensity score models, only model (\ref{correctPS})  is correct, given the true  model (\ref{trtMsimu}). The other models (\ref{wrongPS1})-(\ref{wrongPS3}) misspecified the treatment assignment process by excluding some covariates and interaction terms, or  by using incorrect functional form(s).

\subsection{Evaluation Criteria}

We compared six estimators, where the first four were the standard IPTW  estimators using the individual propensity score models (\ref{wrongPS1})-(\ref{correctPS}), respectively. The final two  estimators were the proposed multiply robust estimators $\widehat\theta_{\E\L}$ and $\widehat\theta_{\m\E\L\C\B}$. Both methods   used all the four models (\ref{wrongPS1})-(\ref{correctPS}) simultaneously. The difference between them was that the final, sixth estimator further incorporated  balance constraints for covariates $X^{(2)}$ and $X^{(3)}$.

Our comparison focused on  the conditional outcome model parameter $\xi=0.5, 1$ or 2. These three cases corresponded to a true causal DDI of  -0.107, -0.170 or -0.231. We considered  a sample size of 2000 and ran 1000 simulations for each parameter configuration. In variance estimation for each estimator, we created 200 bootstrap samples.

We considered three criteria to evaluate the finite sample performance of each estimator. First, we examined the average empirical relative bias across 1000 simulation runs, where the relative bias for an  estimator in one run was defined to be the difference of the estimated and true causal DDIs divided by the true causal DDI. Second, we examined the empirical coverage (in percent), defined to be the percentage of 95\% confidence intervals in 1000 simulation runs that covered the true causal DDI. Third, we examined the empirical standard error across 1000 simulation runs. The first two criteria help assess the statistical consistency and the performance of variance  and 95\% confidence intervals estimation based on bootstrapping. The third criterion is useful to compare the statistical efficiency (or variability) of different estimators.

\subsection{Results}
\subsubsection{Consistency and Robustness}
\begin{figure}[]
\centering
\includegraphics[width=\textwidth]{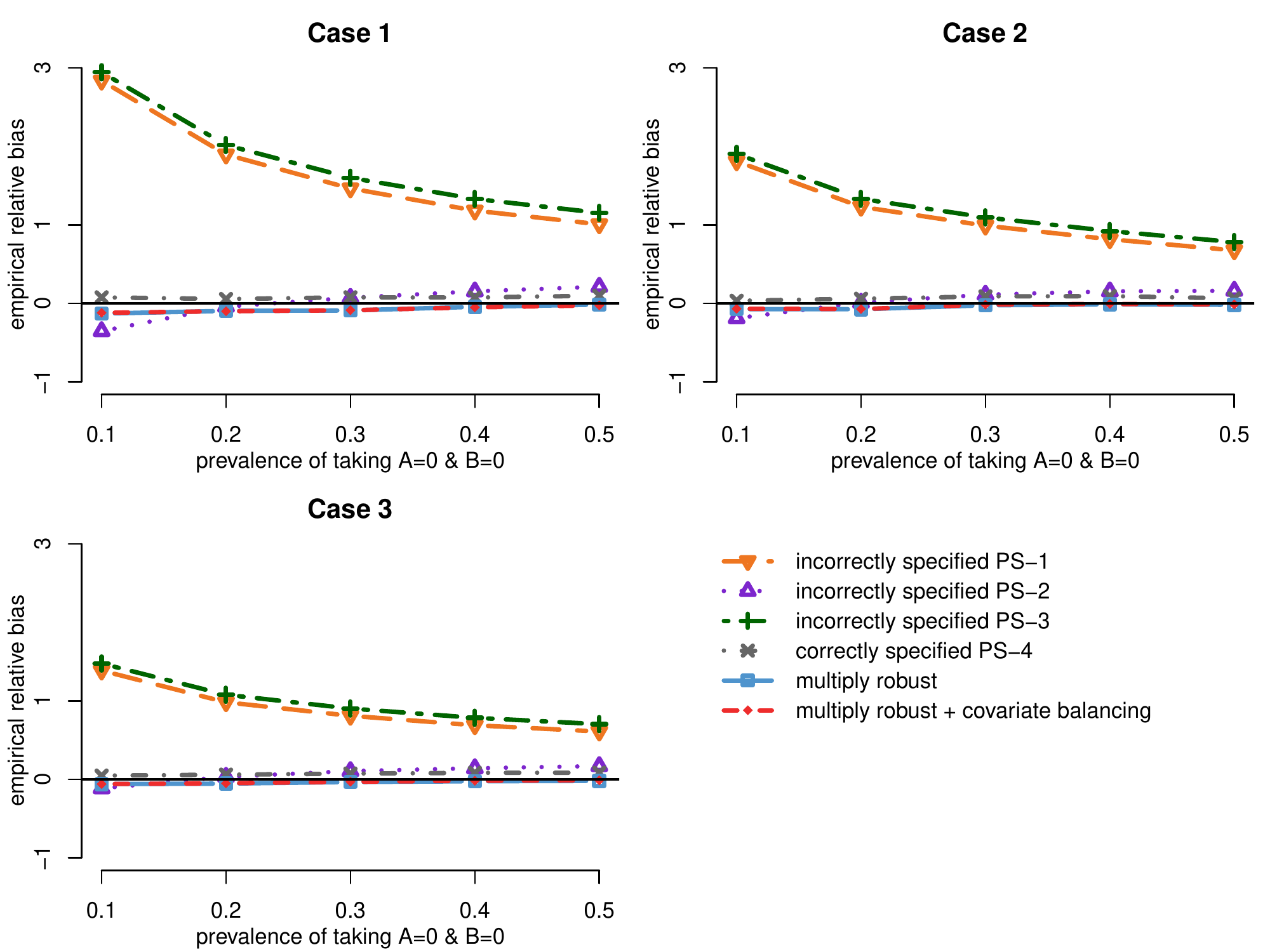}
\caption{Empirical relative bias  with $n=2000$. Incorrectly specified PS-1 to PS-3: IPTW
estimators of the causal DDI using three incorrectly specified propensity score models (\ref{wrongPS1})-(\ref{wrongPS3}), respectively; correctly specified PS-4: IPTW  estimator of the causal DDI using a correctly specified propensity score model (\ref{correctPS}); multiply robust and multiply robust + covariate balancing: the proposed multiply robust estimators using multiple models (\ref{wrongPS1})-(\ref{correctPS}) simultaneously, without and with additional covariate balance constraints, respectively. Cases 1 to 3: the conditional outcome model parameter $\xi=0.5, 1$ or 2,  yielding a true causal DDI of  -0.107, -0.170 or -0.231}
\label{rebias2000}
\end{figure}

Figure \ref{rebias2000} reports the empirical relative bias  for each of the six estimators, under various combinations of causal DDI and treatment prevalence. As expected from their consistency, the IPTW  estimator under a correctly specified propensity score model (\ref{correctPS}) and two proposed multiply robust estimators using  models (\ref{wrongPS1})-(\ref{correctPS}) produced rather small empirical bias. In comparison, due to the use of incorrect  models (\ref{wrongPS1})-(\ref{wrongPS3}), the first three IPTW  estimators led to  biased results. Among the three estimators, the one under the main effects model (\ref{wrongPS2}) produced the least bias; its bias was slightly larger than that of the proposed methods.

\begin{figure}[]
\centering
\includegraphics[width=\textwidth]{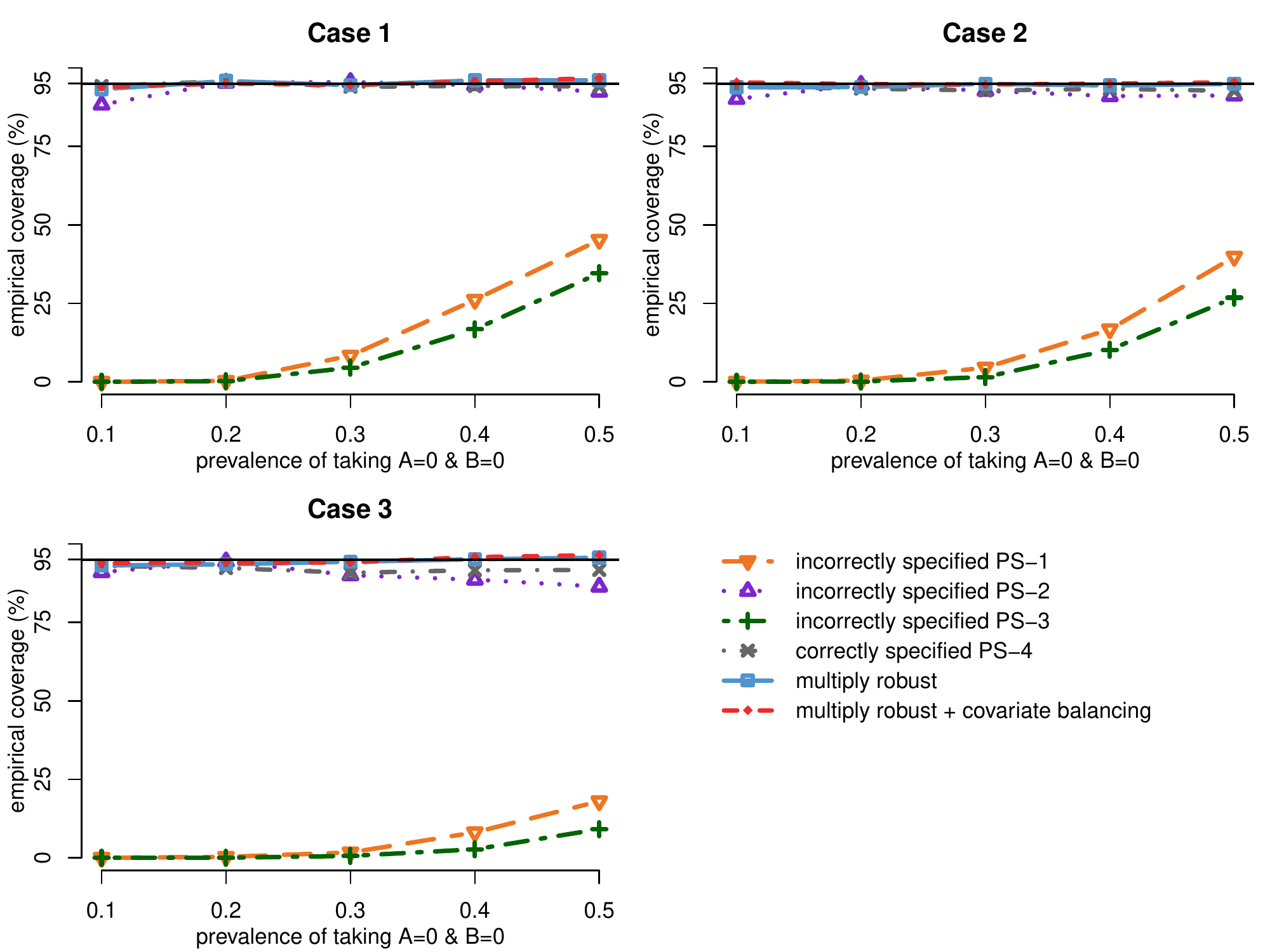}
\caption{Empirical coverage in percent with $n=2000$. Incorrectly specified PS-1 to PS-3: IPTW
estimators of the causal DDI using three incorrectly specified propensity score models (\ref{wrongPS1})-(\ref{wrongPS3}), respectively; correctly specified PS-4: IPTW  estimator of the causal DDI using a correctly specified propensity score model (\ref{correctPS}); multiply robust and multiply robust + covariate balancing: the proposed multiply robust estimators using multiple models (\ref{wrongPS1})-(\ref{correctPS}) simultaneously, without and with additional covariate balance constraints, respectively. Cases 1 to 3: the conditional outcome model parameter $\xi=0.5, 1$ or 2,  yielding a true causal DDI of  -0.107, -0.170 or -0.231}
\label{cp2000}
\end{figure}

Figure \ref{cp2000}  reports the coverage  probability for each of the six estimators. Due to misspecifying the propensity score model, the first three IPTW estimators  yielded  undercovered  95\% confidence intervals.  The IPTW estimator under a correctly specified  model (\ref{correctPS}) and two proposed multiply robust estimators using  models (\ref{wrongPS1})-(\ref{correctPS}) all gave   coverage close to the  nominal  95\%. These results also demonstrated that bootstrapping with 200 replicates performed reasonably well.

\begin{figure}[]
\centering
\includegraphics[width=\textwidth]{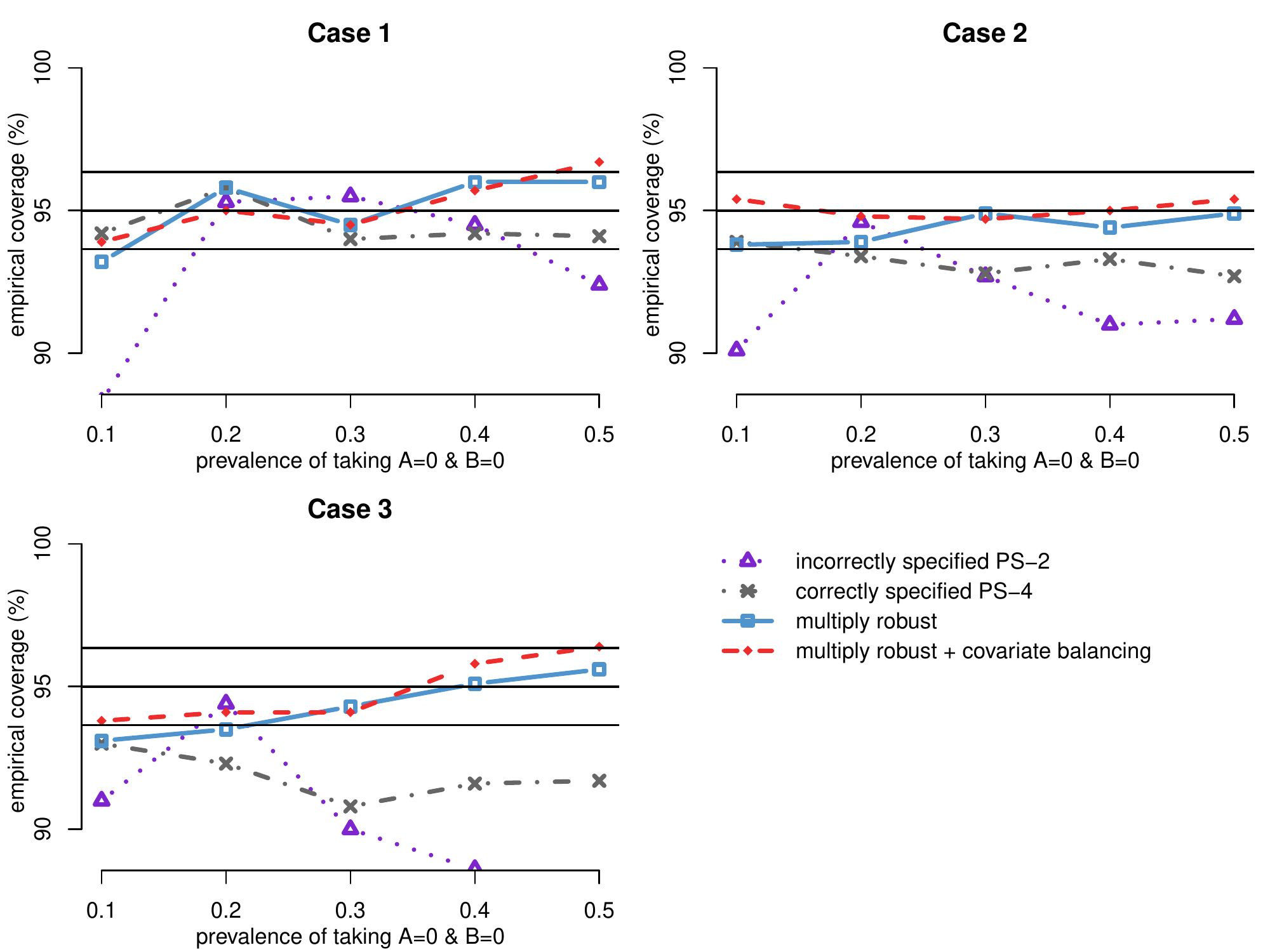}
\caption{A zoom-in version of Figure \ref{cp2000}: empirical coverage in percent with $n=2000$. Three horizontal lines (at 93.65\%, 95\%, and 96.35\%) indicate a plausible range of coverage. Incorrectly specified PS-2: IPTW estimator of the causal DDI using  incorrectly specified propensity score model (\ref{wrongPS2}), i.e., the main effects model; Correctly specified PS-4: IPTW  estimator of the causal DDI using a correctly specified propensity score model (\ref{correctPS}); multiply robust and multiply robust + covariate balancing: the proposed multiply robust estimators using multiple models (\ref{wrongPS1})-(\ref{correctPS}) simultaneously, without and with additional covariate balance constraints, respectively. Cases 1 to 3: the conditional outcome model parameter $\xi=0.5, 1$ or 2,  yielding a true causal DDI  of  -0.107, -0.170 or -0.231}
\label{cpzoom2000}
\end{figure}

To more carefully examine different estimators, Figure \ref{cpzoom2000} is a zoom-in version of Figure \ref{cp2000}. Here, we drew three horizontal lines (at 93.65\%, 95\%, and 96.35\%) to indicate a plausible range of coverage; the approximation formula $0.95\pm 1.96\cdot \sqrt{0.95\times0.05/1000}$  treated coverage status in  each of the 1000 simulation runs as a Bernoulli random variable with success rate 0.95. The empirical coverage rates from both   empirical likelihood-based estimators fluctuated around 95\% and lay within the range of 93.65\% to 96.35\% in most cases. However, we observed slight undercoverage of the IPTW estimator  under a correct  model  (when the prevalence of treatment level $(0,0)$ $>10\%$,  $\xi=1$ or 2 corresponding to a DDI of -0.170 or -0.231). The undercoverage of  the IPTW estimator under the main effects model (\ref{wrongPS2}) is clearly shown. These results showed better finite sample performance of the proposed estimators than the standard IPTW estimator.

From robustness perspective, evidently,  the proposed estimators achieved satisfactory finite sample performance, because they both returned valid results even if involving three misspecified propensity score models.

\begin{center}
{\it [insert Figures \ref{rebias2000}, \ref{cp2000} and \ref{cpzoom2000} here]}
\end{center}

\subsubsection{Efficiency}

We further compared  efficiency performance of two  proposed, multiply robust estimators  with the IPTW estimator under a correct propensity score model (\ref{correctPS}), quantified  by  the empirical standard error.  IPTW estimators under incorrect models were excluded for efficiency  comparison, because for  biased estimators, less variability would only lead to worse coverage.

\begin{figure}[]
\centering
\includegraphics[width=\textwidth]{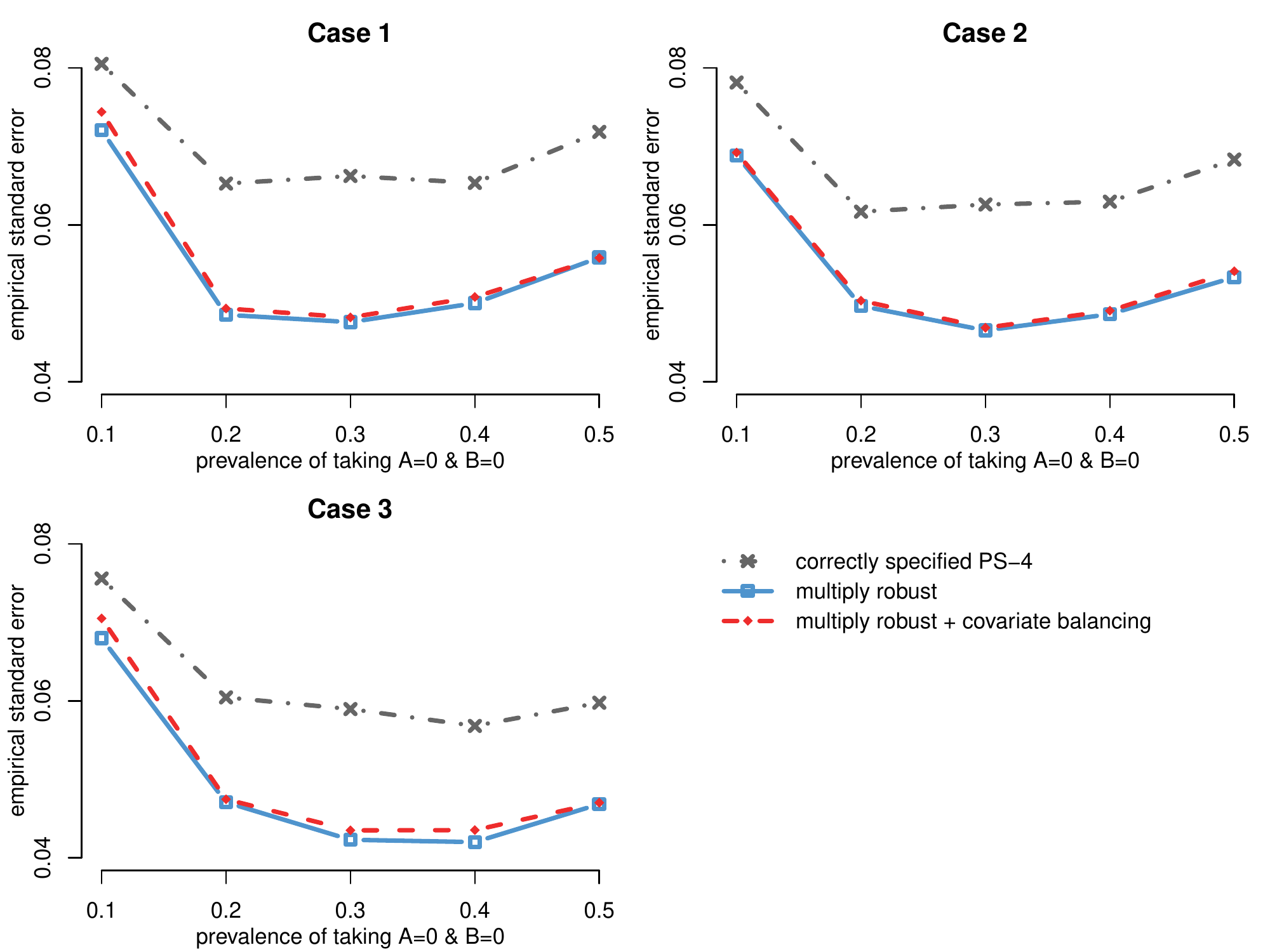}
\caption{Empirical standard error with $n=2000$. Correctly specified PS-4: IPTW  estimator of the causal DDI using a correctly specified propensity score model (\ref{correctPS}); multiply robust and multiply robust + covariate balancing: the proposed multiply robust estimators using multiple models (\ref{wrongPS1})-(\ref{correctPS}) simultaneously, without and with additional covariate balance constraints, respectively. Cases 1 to 3: the conditional outcome model parameter $\xi=0.5, 1$ or 2,  yielding a true causal DDI  of  -0.107, -0.170 or -0.231}
\label{ese2000}
\end{figure}

Figure \ref{ese2000}  displays the results. In all  combinations of causal DDI and treatment prevalence,  on average the first proposed estimator $\widehat\theta_{\E\L}$ produced the smallest standard error, indicating the highest efficiency among three estimators. The other proposed estimator $\widehat\theta_{\m\E\L\C\B}$, from the modified empirical likelihood  approach with  covariate balance, had efficiency  similar to $\widehat\theta_{\E\L}$ and remarkably higher  than the IPTW estimator. These results demonstrated improved efficiency of  the proposed methods relative to the  IPTW estimator.

\begin{center}
{\it [insert Figure \ref{ese2000}  here]}
\end{center}

\section{Simulation Studies: When All Postulated Propensity Score Models Are Wrong}\label{secSimu2}

The consistency of both proposed estimators $\widehat\theta_{\E\L}$ and $\widehat\theta_{\m\E\L\C\B}$  requires a critical condition: at least one candidate model is correct. In this section, we conducted simulations to examine their finite sample performance in the worst scenario when all models were wrong.

\subsection{Setting 1: true model is multinomial logistic}
We conducted simulations  the same as those presented in Section \ref{secSimu1} except that the following model set was used to analyze simulated data:
\begin{equation}
\text{Multinomial logistic}\,\, P(A=a, B=b|\bm X) \sim X^{(1)}\,\,\text{with parameters ${\bm\gamma}^1$},
\label{simu2wrongPS1}
\end{equation}
\begin{equation}
\text{Multinomial logistic}\,\, P(A=a, B=b|\bm X) \sim X^{(4)} + X^{(5)}\,\,\text{with parameters ${\bm\gamma}^2$},
\label{simu2wrongPS2}
\end{equation}
\begin{equation}
\text{Multinomial logistic}\,\, P(A=a, B=b|\bm X) \sim X^{(1)} + X^{(4)} \,\,\text{with parameters ${\bm\gamma}^3$},
\label{simu2wrongPS3}
\end{equation}
\noindent and
\begin{equation}
\text{Multinomial logistic}\,\, P(A=a, B=b|\bm X) \sim X^{(1)} + \exp(X^{(4)})+\exp(X^{(5)}) \,\,\text{with parameters ${\bm\gamma}^4$},
\label{simu2wrongPS4}
\end{equation}
where  ${\bm\gamma}^j$ was the  vector of the parameters in the $j$th propensity score model, for $j=1,\ldots, 4$.

Given the true propensity score model (\ref{trtMsimu}), all these four postulated models are wrong due to excluding some covariates and interaction terms, or  by using incorrect functional form(s).

We compared six estimators. The first four were IPTW  estimators using the individual propensity score models (\ref{simu2wrongPS1})-(\ref{simu2wrongPS4}), respectively. The final two  estimators were the proposed multiply robust estimators $\widehat\theta_{\E\L}$ and $\widehat\theta_{\m\E\L\C\B}$. Both methods   used all  four incorrect models (\ref{simu2wrongPS1})-(\ref{simu2wrongPS4}) simultaneously. Compared with the fifth estimator $\widehat\theta_{\E\L}$, the final, sixth estimator $\widehat\theta_{\m\E\L\C\B}$ further imposed balance constraints for covariates $X^{(2)}$ and $X^{(3)}$. Since models (\ref{simu2wrongPS1})-(\ref{simu2wrongPS4}) all excluded covariates $X^{(2)}$ and $X^{(3)}$, they allowed for more direct evaluation of the estimator $\widehat\theta_{\m\E\L\C\B}$ that aimed to balance the  distributions of covariates $X^{(2)}$ and $X^{(3)}$.

\begin{figure}[]
\centering
\includegraphics[width=\textwidth]{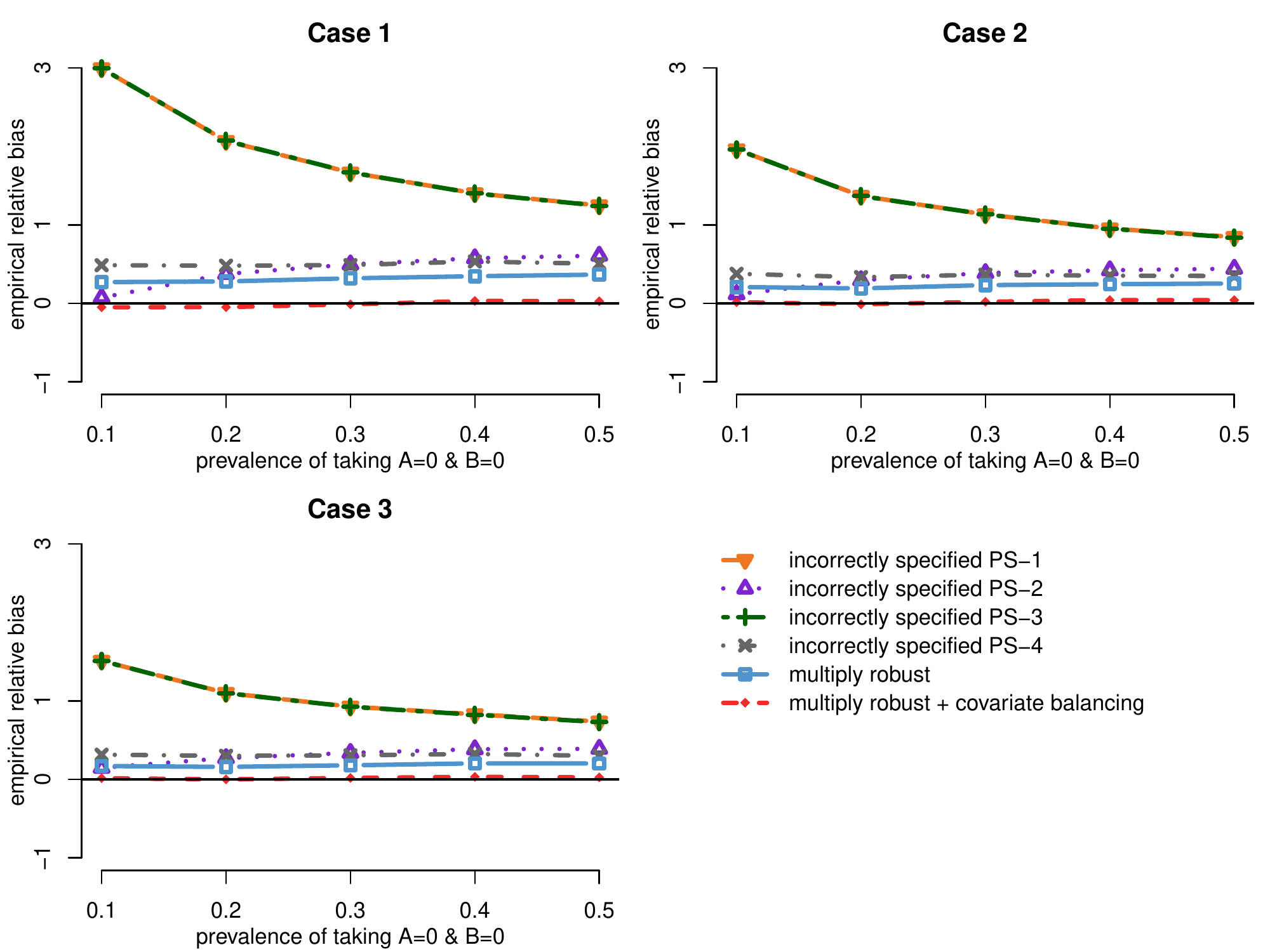}
\caption{When all models are wrong: empirical relative bias  with $n=2000$. Incorrectly specified PS-1 to PS-4: IPTW
estimators of the causal DDI using four incorrectly specified propensity score models (\ref{simu2wrongPS1})-(\ref{simu2wrongPS4}), respectively; multiply robust and multiply robust + covariate balancing: the proposed multiply robust estimators using wrong models (\ref{simu2wrongPS1})-(\ref{simu2wrongPS4}) simultaneously, without and with additional covariate balance constraints, respectively. Cases 1 to 3: the conditional outcome model parameter $\xi=0.5, 1$ or 2,  yielding a true causal DDI of -0.107, -0.170 or -0.231}
\label{rebias2000allwrong}
\end{figure}

\begin{figure}[]
\centering
\includegraphics[width=\textwidth]{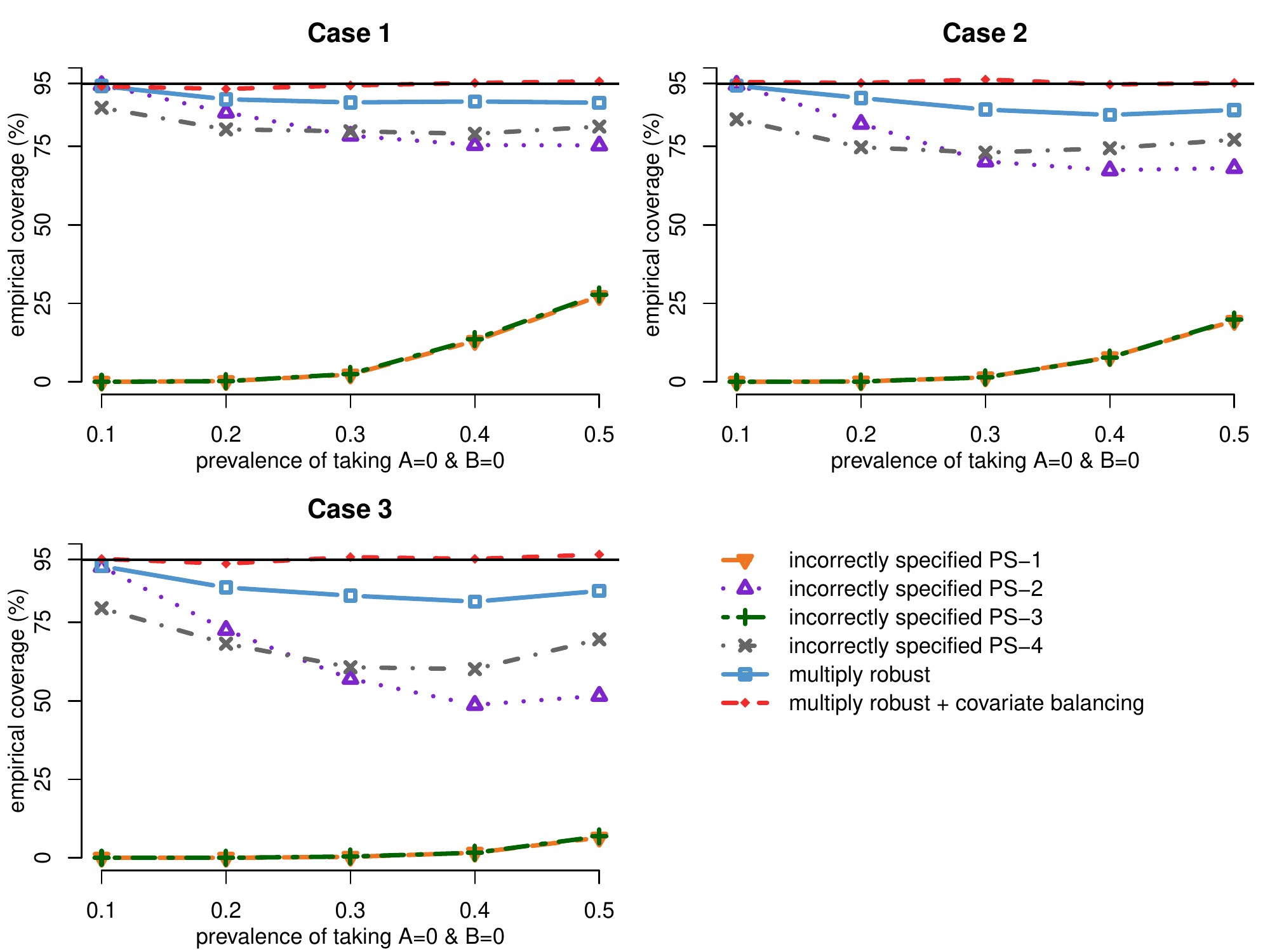}
\caption{When all models are wrong: empirical coverage in percent with $n=2000$. Incorrectly specified PS-1 to PS-4: IPTW estimators of the causal DDI using four incorrectly specified propensity score models (\ref{simu2wrongPS1})-(\ref{simu2wrongPS4}), respectively; multiply robust and multiply robust + covariate balancing: the proposed multiply robust estimators using wrong models (\ref{simu2wrongPS1})-(\ref{simu2wrongPS4}) simultaneously, without and with additional covariate balance constraints, respectively. Cases 1 to 3: the conditional outcome model parameter $\xi=0.5, 1$ or 2,  yielding a true causal DDI of -0.107, -0.170 or -0.231}
\label{cp2000allwrong}
\end{figure}

For each of the six estimators, Figures \ref{rebias2000allwrong} and \ref{cp2000allwrong}  summarize the empirical relative bias and coverage probability, respectively, under various combinations of causal DDI and treatment prevalence. Four IPTW estimators using  individual, wrong models  (\ref{simu2wrongPS1})-(\ref{simu2wrongPS4}) all led to severely biased results and undercoverage  due to  misspecifying the treatment decision process. The proposed estimator $\widehat\theta_{\E\L}$ had  slightly better performance compared to the least-biased IPTW estimator (among all IPTW estimators). Performance of the  proposed estimator $\widehat\theta_{\m\E\L\C\B}$ was promising. It produced very small empirical bias and  satisfactory empirical coverage close to the  nominal  95\%. The better performance of $\widehat\theta_{\m\E\L\C\B}$ compared to $\widehat\theta_{\E\L}$ was probably due to the use of additional balance conditions for covariates $X^{(2)}$ and $X^{(3)}$, given that information on $X^{(2)}$ and $X^{(3)}$ was not used by any of the models (\ref{simu2wrongPS1})-(\ref{simu2wrongPS4}). We note that  such simulation design was intended to examine the  nonparametric balance constraints for $\widehat\theta_{\m\E\L\C\B}$. The results do not necessarily suggest that  $\widehat\theta_{\m\E\L\C\B}$ generally outperforms $\widehat\theta_{\E\L}$. The reason is, if information on $X^{(2)}$ and $X^{(3)}$  was leveraged by $\widehat\theta_{\m\E\L\C\B}$,   such information could also be leveraged by $\widehat\theta_{\E\L}$ too, by including $X^{(2)}$ and $X^{(3)}$ in propensity score model(s).

\begin{figure}[]
\centering
\includegraphics[width=\textwidth]{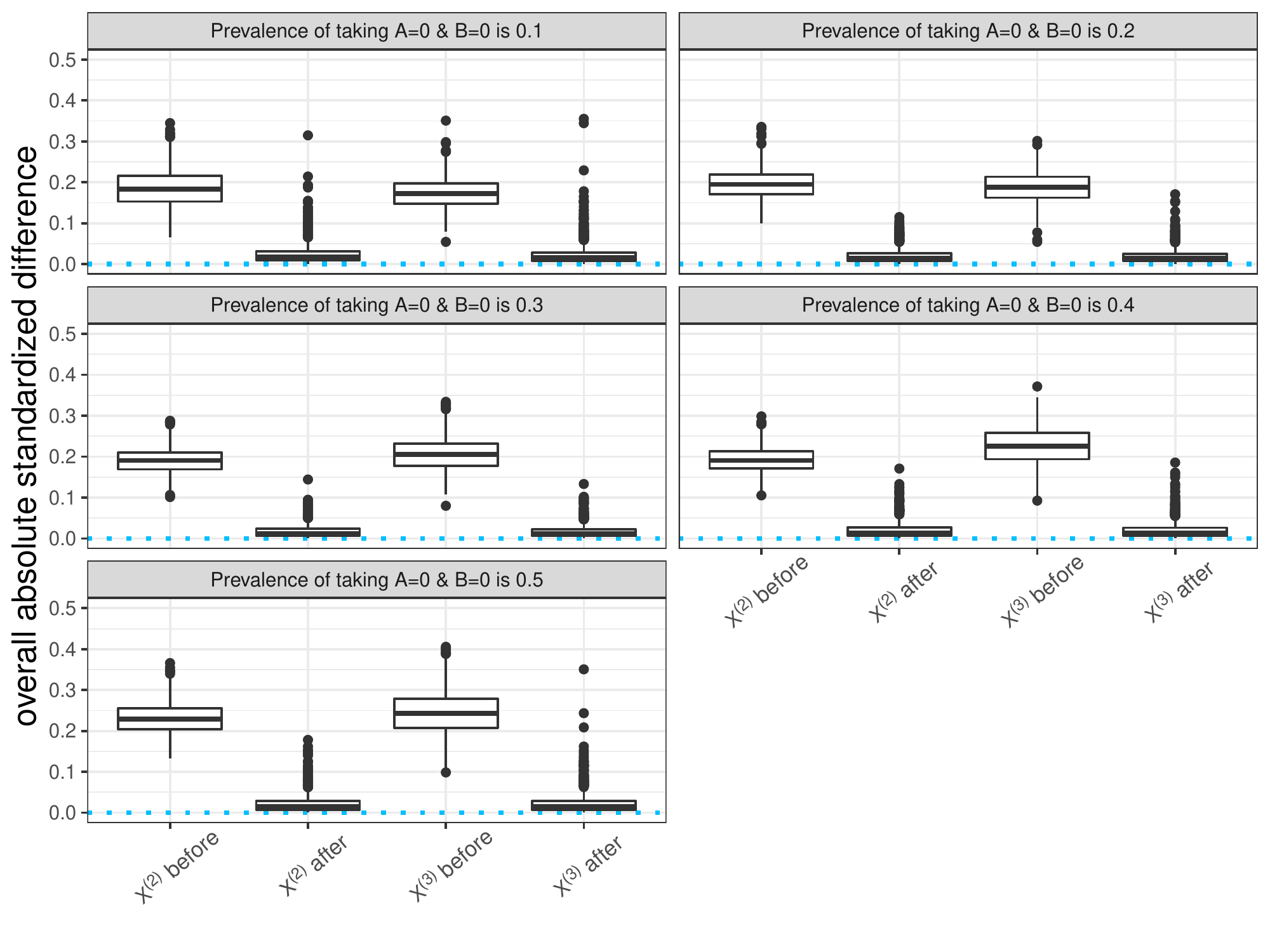}
\caption{When all models are wrong: boxplot of overall balance (defined as the maximum absolute  standardized difference across four drug pairs) of $X^{(2)}$ and $X^{(3)}$ across 1000 simulation runs, before and after weighting data using the proposed multiply robust estimator with balance constraints on $X^{(2)}$ and $X^{(3)}$. The dotted horizontal line represents zero imbalance}
\label{balanceBoxplot}
\end{figure}

To directly examine the performance of balance conditions for $\widehat\theta_{\m\E\L\C\B}$,  Figure \ref{balanceBoxplot} displays boxplot of overall balance (i.e., the maximum absolute  standardized difference across four drug pairs) of $X^{(2)}$ and $X^{(3)}$ across 1000 simulation runs, before and after weighting data using the proposed multiply robust estimator with balance constraints for $X^{(2)}$ and $X^{(3)}$. The results
showed negligible imbalance in the majority of runs under various treatment prevalence. Compared to balance results before weighting, results after weighting were significantly improved.
Given that all postulated models did not involve $X^{(2)}$ and $X^{(3)}$, it is reasonable to attribute  good covariate balancing to the balance constraints for $X^{(2)}$ and $X^{(3)}$.

\begin{center}
{\it [insert Figures \ref{rebias2000allwrong}, \ref{cp2000allwrong} and \ref{balanceBoxplot} here]}
\end{center}

\subsection{Setting 2: true model is not multinomial logistic}

We conducted simulations  the same as those presented in Section \ref{secSimu1} except that a  non-multinomial logistic model served as the true propensity score model.

We first generated two continuous random variables  $Z_1=Q_{10}+Z_0$ and $Z_2=Q_{01}-Z_0$, where $Q_{10}$ and $Q_{01}$ were defined as in Section \ref{secSimu1}  with $\gamma_0=0$, and $Z_0$ followed the standard normal distribution. Then, we simulated treatment indicators as below:
\begin{equation}
 (A, B) = \left\{
\begin{array}{ll}
      (1,1), & \text{if}\,\,  Z_1\ge c, Z_2\ge c \\
      (1,0), & \text{if}\,\,  Z_1\ge c, Z_2 < c \\
     (0,1), & \text{if}\,\,  Z_1 < c, Z_2\ge c \\
     (0,0), & \text{if}\,\,  Z_1 < c, Z_2 < c \\
\end{array} 
\right.
\label{psWrg}
\end{equation}
where  constant $c$ was chosen to yield desired prevalence of treatment  level $(0,0)$ of  10\%, 20\%, 30\%, 40\% or 50\%.

Following Section \ref{secSimu1}, we postulated multinomial logistic models (\ref{wrongPS1})-(\ref{correctPS}) to analyze simulated data. Since the true model (\ref{psWrg}) is not multinomial logistic, all postulated models are incorrect.

\begin{figure}[]
\centering
\includegraphics[width=\textwidth]{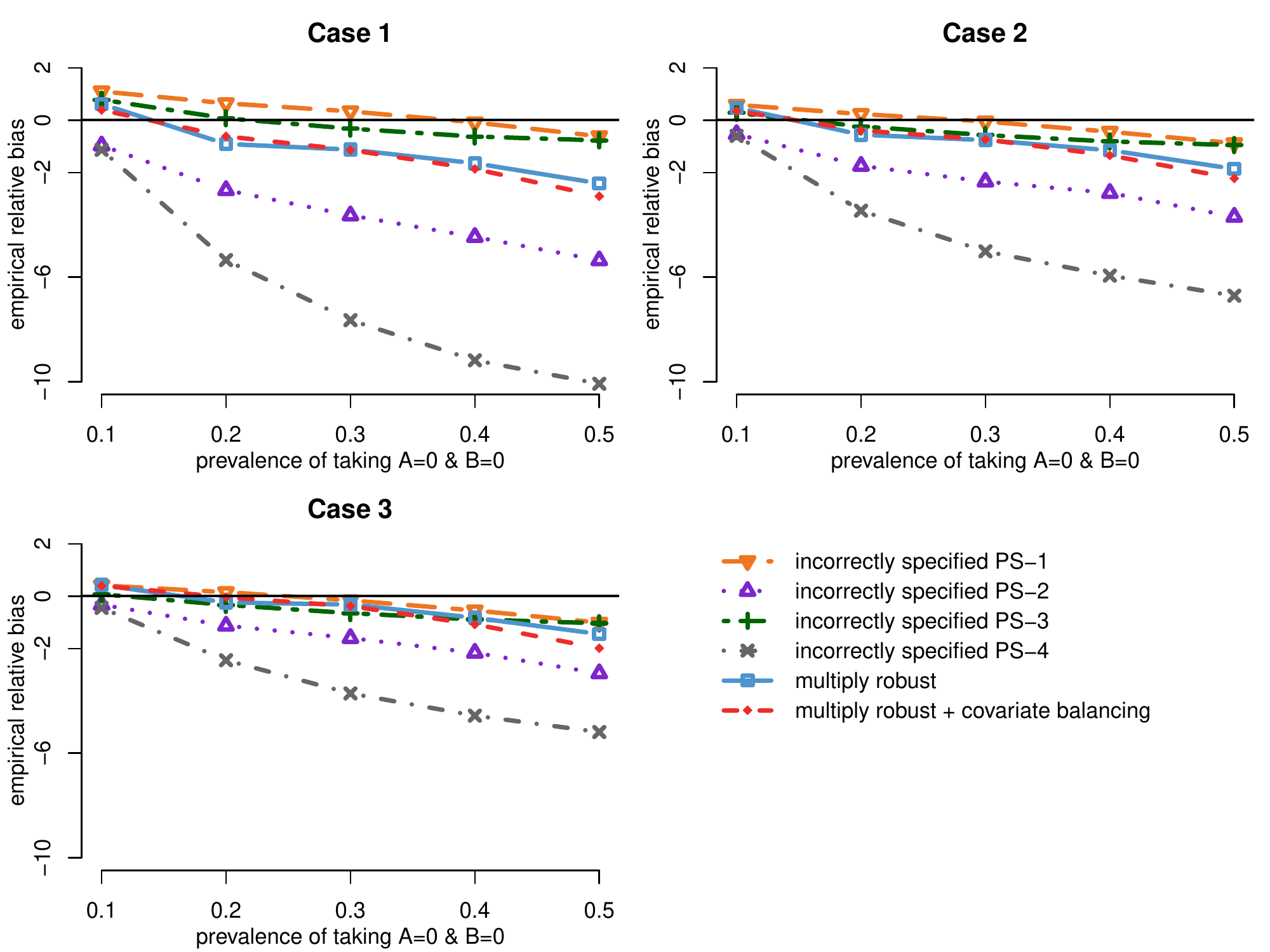}
\caption{When all models are wrong and the true model (\ref{psWrg}) is not multinomial logistic: empirical relative bias  with $n=2000$. Incorrectly specified PS-1 to PS-4: IPTW
estimators of the causal DDI using four incorrectly specified propensity score models (\ref{wrongPS1})-(\ref{correctPS}), respectively; multiply robust and multiply robust + covariate balancing: the proposed multiply robust estimators using wrong models (\ref{wrongPS1})-(\ref{correctPS}) simultaneously, without and with additional covariate balance constraints, respectively. Cases 1 to 3: the conditional outcome model parameter $\xi=0.5, 1$ or 2,  yielding a true causal DDI of  -0.107, -0.170 or -0.231}
\label{rebiasWrg}
\end{figure}

Figure \ref{rebiasWrg} reports the results. All models produced severely biased results due to misspecification. Of note, IPTW estimators under fuller models (\ref{wrongPS2}) and (\ref{correctPS}) had  even worse performance compared with  models (\ref{wrongPS1}) and (\ref{wrongPS3}), demonstrating that adding more covariates into an existing propensity score model can paradoxically increase bias. The  two proposed estimators outperformed IPTW estimators under models (\ref{wrongPS2}) and (\ref{correctPS}) but not IPTW estimators under models (\ref{wrongPS1}) and (\ref{wrongPS3}).

\begin{center}
{\it [insert Figure \ref{rebiasWrg} here]}
\end{center}

\section{Application to  Kidney Data}\label{secAppl}

We applied the proposed methods from Sections \ref{secProp1} and \ref{secProp2}, and  the standard IPTW method  from Section \ref{secIPTW}  in order to analyze data from a pharmacoepidemiologic study on possible  impact of renin-angiotensin system inhibitors (RAS-I) on the comparative nephrotoxicity of nonsteroidal anti-inflammatory drugs (NSAIDS) and opioids. \citep{Miano2020} 

The data set  included 27,741 patients who were age $\ge 18$  years and admitted to one of four hospitals within the University of Pennsylvania Health System from January 1, 2004 to June 30, 2017. NSAIDS and RAS-I were considered the possible object and precipitant, respectively. Lacking effects on kidney function, oxycodone and amlodipine were considered comparator drugs. Specifically, the first treatment indicator $A$ was  set to 1 for NSAIDS  and 0 for oxycodone. The second treatment indicator $B$ was  set to 1 for RAS-I  and 0 for amlodipine. Eligible patients had to maintain at least 24 hours of concomitant treatment with a drug pair of interest.  Among these patients, 4,250 (15.3\%) were treated with the  NSAIDS+RAS-I drug pair, 1,181 (4.3\%)  the NSAIDS+amlodipine drug pair, 17,610 (63.5\%) the oxycodone+RAS-I  drug pair, and 4,700 (16.9\%)  the oxycodone+amlodipine drug pair. The outcome was a binary indicator for acute kidney injury (AKI) during the first 14 days of follow-up, defined using Kidney Disease Improving Global Outcomes (KDIGO) creatinine and dialysis criteria.  A total of 2,486  patients (9.0\%) were observed to experience the AKI  incident. Our goal is to estimate the causal DDI between NSAIDS and RAS-I, defined in (\ref{ddi}) on the additive scale, i.e., a difference of risk differences.

We built a set of multinomial logistic propensity score models guided by clinical knowledge and prior literature. We began with the first model (PS-1) that included baseline covariates that have clearly been identified as risk factors for the outcome of AKI in previous studies. They fell into six categories (a) demographics, including age and body mass index (BMI); (b) hospital admission characteristics, including admission type,  treatment in an ICU at the time of cohort entry, and postoperative status; (c) comorbidities, including chronic kidney disease, heart failure, myocardial infarction, hypertensions, other cardiac arrhythmia, liver disease, and diabetes mellitus; (d) kidney function, including baseline glomerular filtration rate, and history of prior AKI; (e) labs, including baseline hemoglobin concentration, and baseline chloride concentration; (f) medications, including loop diuretics, hydrochlorothiazide,  vancomycin,  trimethoprim/sulfamethoxazole,  antibiotic nephrotoxins, and other miscellaneous nephrotoxins.

The second model (PS-2) included variables that have clearly been identified as risk factors for the outcome of AKI, with additional variables that may be associated with treatment decisions, but are less well established as risk factors. Specifically, we added variables to PS-1, which fell into categories (a) demographics, including sex and race; (c) comorbidities, including atrial fibrillation, valvular disease, pulmonary circulation disease, chronic pulmonary disease, cancer, and obstructive sleep apnea.

The third model (PS-3) included variables from PS-2 plus variables that may be associated with treatment decisions and also may function as overall markers of illness severity. Specifically, we added variables to PS-2, which fell into categories (b)  hospital admission characteristics, including hospital where the patient was treated, and duration of hospitalization prior to cohort entry, in days; (c) comorbidities, including unexplained weight loss, fluid or electrolyte disorder, and human immunodeficiency virus infection; (e) labs, including baseline white blood cell count, baseline platelet count, and baseline potassium concentration; f) medications, including beta-blockers, alpha-beta blockers, hydralazine, other miscellaneous antihypertensive medications, broad-spectrum gram negative antibiotics, and narrow spectrum gram negative antibiotics.

The fourth model (PS-4) included variables from PS-3 plus interaction terms between admission type and the following variables: age, chronic kidney disease, heart failure, baseline hemoglobin concentration, and baseline platelet count. To further capture nonlinear terms, the fifth, final model (PS-5) included variables from PS-4 plus quadratic terms for the following variables: body mass index, baseline glomerular filtration rate, baseline hemoglobin concentration, and baseline chloride concentration.

We compared seven estimators of the causal DDI, quantified by comparing the difference in AKI risk between NSAID versus oxycodone in patients treated with RAS-I to the difference in AKI risk between NSAID versus oxycodone in patients treated with amlodipine. The first, second, third, fourth and fifth estimators  were the standard  IPTW estimators using individual models PS-1, PS-2, PS-3, PS-4 and PS-5, respectively. The sixth and seventh estimators were obtained by applying the  proposed multiply robust methods described in Sections \ref{secProp1} and  \ref{secProp2}, respectively. Both estimators used  all five  propensity score models simultaneously, motivated by an understanding that the true propensity score model was unknown; in this reality, the fifth, fullest model PS-5 is not necessarily the best-performing one among all five models. Compared to the sixth estimator, the seventh estimator imposed additional balance constraints for age and chronic kidney disease;  these two covariates were selected for illustration purpose given that they were both known risk factors and were the only two covariates having  overall absolute standardized difference >15\%  before and after weighting the data sample using IPTW under the fullest model PS-5. For each of these seven estimators, we bootstrapped the data set 200 times to estimate the variance and 95\% confidence interval.

Table \ref{tb1} summarizes the analysis results. The five IPTW estimators produced larger (further from the null) causal DDI estimates than two proposed methods.  PS-3 and PS-4 produced IPTW-based DDI estimates  greater than twice the size of  estimates from the proposed methods. The estimator based on PS-1 produced slightly smaller estimates than  that based on PS-5, but the difference was not large. This implied that once  the key risk factors (that have clearly been identified as risk factors for the outcome of AKI in previous studies) have been adjusted for, further including  possible risk factors, other surrogate markers of overall severity of illness, interaction terms and non-linear terms, did not make a remarkable difference. The estimates based on PS-3 and PS-4  were similar, and were larger than results from PS-1, PS-2 and PS-5.  Although DDI estimates differed from method to method, the differences between methods were small on the additive scale. All methods produced 95\% confidence intervals for the causal DDI  that included 0, suggesting no evidence of a statistically significant synergistic interaction  at the 5\%  level. The point estimates were close to the null value zero (e.g., 0.200$\%$ and 0.146$\%$ from two proposed methods), implying a small excess risk attributed to RAS-I when assessing the NSAID-AKI risk.  These results were consistent with prior studies.  \citep{Miano2020} The estimate based on PS-1 (the simplest model) had the least standard error, and  the estimate based on PS-5 (the fullest model) had the largest standard error. Two proposed methods produced estimates that had less standard errors than all IPTW estimates except for the estimate based on PS-1. The last estimate, based on modified empirical likelihood procedures, had standard error slightly higher than but comparable to the estimate based on PS-1.

\begin{center}
{\it [insert Table \ref{tb1}   here]}
\end{center}

\begin{figure}[]
\centering
\includegraphics[width=\textwidth]{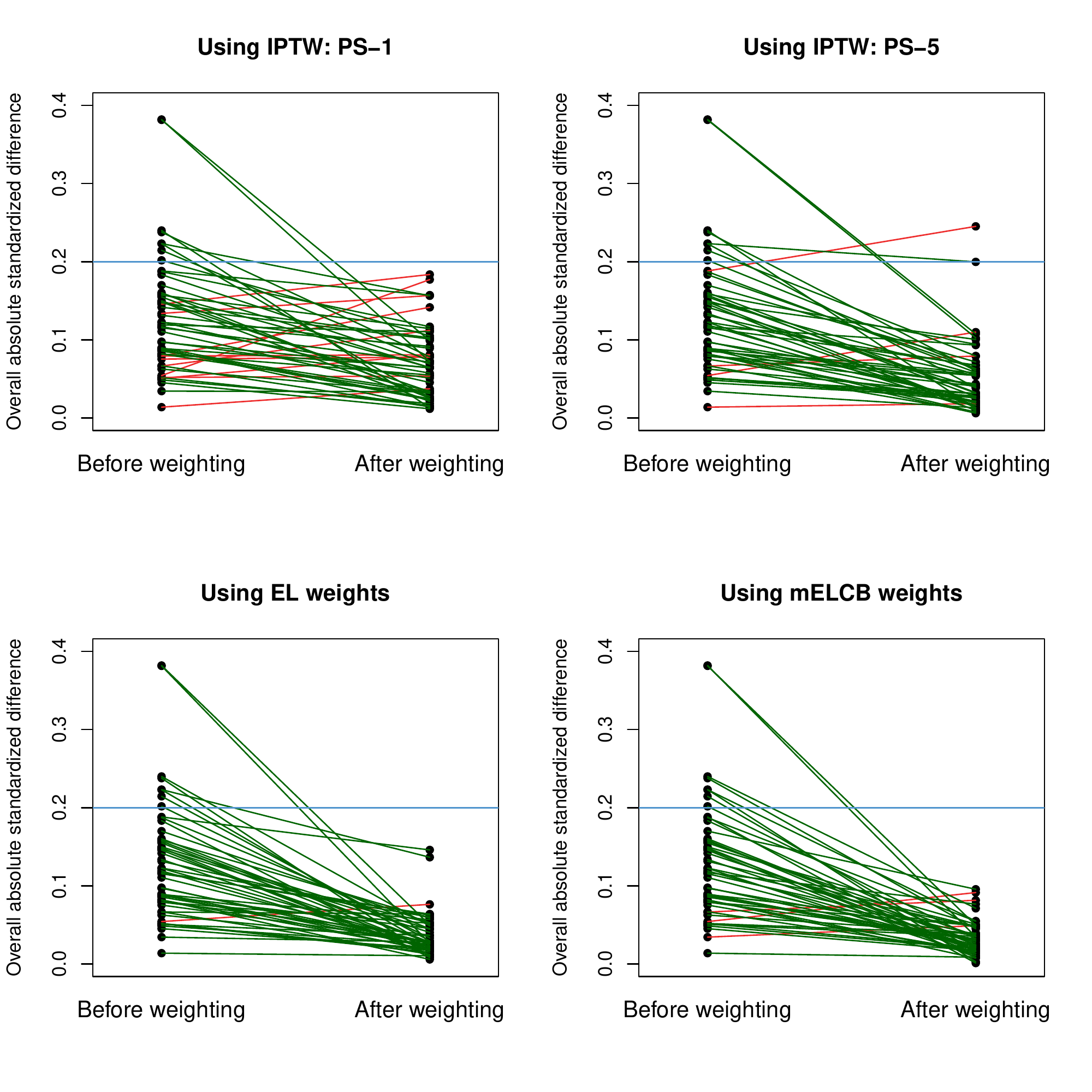}
\caption{Overall absolute  standardized differences of baseline covariates before and after weighting the kidney data sample using various methods: i) IPTW estimator based on the simplest model PS-1, ii) IPTW estimator based on the fullest model PS-5, iii)  the first proposed multiply robust estimator (EL) using models PS-1 to PS-5, and   iv) the  second proposed multiply robust estimator (mELCB)  with balance constraints on age and chronic kidney disease}
\label{overallbalance}
\end{figure}

To assess covariate balance, Figure \ref{overallbalance} visualizes the overall absolute standardized differences across drug pairs for 52 baseline covariates (including dummy variables for categorical data),  before and after weighting the sample using the following methods: i) IPTW under the simplest model PS-1, ii) IPTW under the fullest model PS-5, iii)  the proposed empirical likelihood method using models PS-1 to PS-5, and  iv) the proposed modified empirical likelihood  method with additional balance constraints on age and chronic kidney disease. We drew a horizontal line to represent a cutoff value of 0.2, beyond which a balance result might be problematic. \citep{McCaffrey2013}  Without any confounding adjustment (i.e., before weighting), several baseline covariates had  remarkably imbalanced distributions. In particular, 8 covariates had overall standardized mean differences $>0.2$ for the unweighted sample. On contrary, two proposed empirical  likelihood weighting methods were found to well balance all measured covariates  in the weighted sample; no covariate had an overall standardized mean difference $>0.2$. If we considered a more conservative cutoff value of 0.1, then 29 covariates had overall standardized mean differences $>0.1$ for the unweighted sample. In comparison,  2 covariates for the empirical likelihood weighting method (without additional covariate balance constraints), and no covariate for the modified empirical likelihood weighting method (with additional  balance constraints),  had an overall standardized mean difference $>0.1$.

In terms of covariate balancing, the proposed methods appeared to outperform the IPTW estimators based on PS-1 and PS-5, which led to 13 and 6 covariates having an overall standardized mean difference $>0.1$ for the weighted sample. Comparing two IPTW estimators, the more complex model PS-5 performed better than the simpler model PS-1.

We further examined  covariate balance performance of  the  proposed modified empirical likelihood method that directly balanced the distributions of age and chronic kidney disease for each drug pair. Table \ref{tb2}  reports absolute  standardized differences of age and chronic kidney disease in each of the four drug pairs. As expected, the modified empirical likelihood method produced negligible standardized mean differences for both covariates.

\begin{center}
{\it [insert Table \ref{tb2} and Figure \ref{overallbalance}  here]}
\end{center}

\section{Discussion}\label{secDiscuss}

In this paper, we have proposed two multiply robust methods for estimating the causal DDIs by making use of a set of propensity score models instead of just one. The second method further reduces model dependence by adding constraints to directly balance the distributions of user-specified covariates. Both methods guarantee statistical  consistency, as long as one of the candidate models is correctly specified (and in this case, there is no need to identify which model is correct). Compared to the standard IPTW  method that relies on a single propensity score model, the proposed methods provide increased protection against model misspecification. Although we place the development in two-drug DDI context, the methodology can be directly applied to general multi-valued treatment case, and to other applications  such as  interactions between genetic and environmental exposures. \citep{vanderweele2014}

Our simulations demonstrated severe bias and undercoverage  due to misspecifying a propensity score model when using the IPTW method.  Simulation results also  confirmed satisfactory finite sample performance of the proposed methods in terms of robustness. Moreover,  our methods produced a smaller standard error compared to the IPTW  method under a correct model, suggesting an efficiency gain without losing consistency.  This is consistent with numerical findings in a prior study. \citep{Shu2021} Similarly, in weight calibration context, Han \cite{Han2018} illustrated that including incorrect models can  enhance  efficiency  when a correct propensity score model is specified, by examining the projection residual. In simulations settings where all postulated models were severely misspecified, both IPTW and the proposed methods can lead to biased results, suggesting the need to include good candidate model(s) when implementing the proposed methods.

In theory, the proposed methods allow for any finite number of candidate models, and more models  would naturally offer a better chance of including a correct one. However, numerically, having too many models/constraints may result in challenges such as collinearity. \citep{Han2014a, Han2014b, Han2016}  It is advisable to only include reasonable models that are not too similar with each other. In our limited experiences, postulating $J=5$ models or so does not lead to numerical challenges. Such number of models might be sufficient in many applications. The hope is that after careful considerations with subject matter knowledge,  a short list of models will be generated  to form the model set. For example, our kidney data analysis involves five models, where Model 1 includes key risk factors only, Model 2 = Model 1 + possible risk factors, Model 3 = Model 2 + other surrogate markers of the outcome, Model 4 = Model 3 + interaction terms, and Model 5 = Model 4 + non-linear terms. This 5-model strategy may or may not be the most meaningful in other studies. We need more discussion on actionable model specification strategies to facilitate the implementation of multiply robust estimation.

As for what covariates to impose balance constraints when implementing the modified empirical likelihood method, there are at least two strategies. The first is to specify  {\it a priori}, important covariates using subject matter knowledge prior to the collection and analysis of data. The second is to  select covariates that are severely imbalanced in the unweighted data or are suboptimally balanced in another weighted analysis (e.g., the original empirical likelihood method without additional  balance constraints). The second strategy may serve  as a means to handle residual imbalance. Evaluating different strategies is beyond the scope of the current work. More generally, the proposed methods might be employed in combination with matching for improved covariate balance. When matching alone does not well balance all covariates, one could apply empirical likelihood weighting presented here to the matched sample.

The distance between two proposed estimators (when postulating  the same model set) might be used to detect possible model misspecification. The reason is that if the model set contains a correctly specified propensity score model,  the resulting estimators from both methods are consistent, and their difference should converge to zero as sample size grows. Therefore, a significant discrepancy between these two estimates may indicate that none of the candidate models is correct.  Formal tests using this idea were developed in various contexts.  \citep{Hausman1978,ShuHe2021}  It is useful to study specification tests in our multiple-model context, although non-significant results cannot confirm the inclusion of correct model(s).

The current development does not use any outcome model, making it easy to examine different outcome variables; thus there is no need to re-calculate the empirical likelihood weights for DDIs with respect to a new outcome. Future work will  further include multiple conditional outcome models and evaluate its statistical performance for the estimation of causal DDIs.

\section*{ACKNOWLEDGMENTS}
The authors thank the reviewers for their constructive comments which led to an improved version of this paper. This work was funded by NIH grants K08DK124658, R01DA048001, R01AG064589, and R01AG025152.

\section*{DATA AVAILABILITY STATEMENT}
The real-world kidney data derived from electronic medical records from the University of Pennsylvania Health System are currently not available for public sharing.

\bibliography{MRref}

\newpage

\begin{table}[!t]
\centering
\tabcolsep20pt
\caption{Analysis results of the real-world kidney data estimating the causal DDI between NSAIDS and RAS-I. The  DDI was quantified by comparing the difference in AKI risk between NSAID vs oxycodone in patients treated with RAS-I to the difference in AKI risk between NSAID vs oxycodone in patients treated with amlodipine}
\begin{threeparttable}
\begin{tabular}{ccccc}
\toprule
Method & Model        & DDI      & Standard error & $95\%$ Confidence interval \\
\midrule
IPTW   & PS-1         & 0.259$\%$  & 0.0144         & $(-2.55$\%$, 3.07$\%$)$        \\
       & PS-2         & 0.354$\%$  & 0.0153         & $(-2.65$\%$, 3.36$\%$)$        \\
       & PS-3         & 0.524$\%$  & 0.0150         & $(-2.41$\%$, 3.46$\%$)$        \\
       & PS-4         & 0.509$\%$  & 0.0153         & $(-2.49$\%$, 3.51$\%$)$        \\
       & PS-5         & 0.324$\%$  & 0.0159         & $(-2.80$\%$, 3.44$\%$)$        \\\midrule
EL     & PS-1 to PS-5    & 0.200$\%$  & 0.0150         & $(-2.74$\%$, 3.14$\%$)$        \\
mELCB  & PS-1 to PS-5& 0.146$\%$ & 0.0146         & $(-2.71$\%$, 3.00$\%$)$   \\
\bottomrule    
\end{tabular} {The DDI column represents the difference-in-difference estimate;\\
PS-1: multinomial logistic propensity score model including key risk factors; \\
PS-2: multinomial logistic propensity score model including key risk factors and possible risk factors;\\
PS-3: multinomial logistic propensity score model including key risk factors, possible risk factors, and other surrogate markers of overall severity of illness; \\
PS-4: multinomial logistic propensity score model including key risk factors, possible risk factors,  other surrogate markers of overall severity of illness, and interaction terms;\\
PS-5: multinomial logistic propensity score model including key risk factors, possible risk factors,  other surrogate markers of overall severity of illness, interaction terms, and non-linear terms;\\
EL: the first proposed multiply robust estimator using all five propensity score models simultaneously;\\
mELCB: the second proposed multiply robust estimator using all five propensity score models simultaneously, with additional balance constraints on age and chronic kidney disease;\\
(see text for the list of selected covariates in each model)}
\end{threeparttable}
\label{tb1}
\end{table}

\begin{table}[!t]
\centering
\tabcolsep4.5pt
\caption{Absolute  standardized differences of age and chronic kidney disease (ckd)  in each of the four drug pairs, before and after weighting the kidney data sample using various methods: i) IPTW estimator based on the simplest model PS-1, ii) IPTW estimator based on the fullest model PS-5, iii)  the first proposed multiply robust estimator (EL) using models PS-1 to PS-5, and   iv) the  second proposed multiply robust estimator (mELCB)  with balance constraints on age and ckd}
\begin{threeparttable}\begin{tabular}{ccccccccccccc}
\toprule
           & \multicolumn{2}{c}{NSAIDS+RAS-I}          &  & \multicolumn{2}{c}{NSAIDS+Amlodipine} &  & \multicolumn{2}{c}{Oxycodone+RAS-I}       &  & \multicolumn{2}{c}{Oxycodone+Amlodipine} &    \\ \cline{2-3} \cline{5-6} \cline{8-9}\cline{11-12} \
Method     & age              & ckd &  & age       & ckd    &  & age              & ckd&  & age        & ckd      &    \\ \midrule
Unadjusted & 0.220            & 0.188$^\ddagger$                  &  & 0.223$^\dagger$     & 0.169                     &  & 0.043            & 0.022                  &  & 0.095      & 0.129                       &    \\
IPTW PS-1  & 0.005            & 0.020                  &  & 0.156$^\dagger$     & 0.157$^\ddagger$                     &  & 0.005            & 0.003                  &  & 0.047      & 0.014                       &   \\
IPTW PS-5  & 0.009            & 0.015                  &  & 0.200$^\dagger$     & 0.245$^\ddagger$                     &  & 0.005            & 0.005                  &  & 0.048      & 0.029                       &    \\
EL         & 0.016            & 0.005                  &  & 0.137$^\dagger$     & 0.146$^\ddagger$                     &  & 0.005            & 0.005                  &  & 0.051      & 0.027                       &    \\
mELCB      & \textless{}0.001 & \textless{}0.001       &  & 0.002     & \textless{}0.001          &  & \textless{}0.001 & \textless{}0.001       &  & 0.003$^\dagger$      & 0.001$^\ddagger$                       &  \\
\bottomrule    
\end{tabular} {$^\dagger$ indicates the overall balance of age for each method, defined as the maximum absolute  standardized difference across four drug pairs; \\
$^\ddagger$ indicates the overall balance of  ckd  for each method, defined as the maximum absolute  standardized difference  across four drug pairs}
\end{threeparttable}
\label{tb2}
\end{table}

\end{document}